\documentclass[epj]{svjour}

\usepackage{graphicx}
\usepackage{amsmath}

\begin{document}

\title{Magnetic reordering in the vicinity of a \\
  ferromagnetic/antiferromagnetic interface}

\author{P. J. Jensen\inst{1} \thanks{Corresponding author}
\and H. Dreyss\'e\inst{2} \and Miguel Kiwi\inst{3}}
\institute{Institut f\"ur Theoretische Physik,
Freie Universit\"at Berlin, Arnimallee 14, D-14195 Berlin, Germany 
\and 
IPCMS -- GEMME, Universit\'e Louis Pasteur, 23, rue du Loess,
F-67037 Strasbourg, France 
\and
Facultad de F\'{\i}sica, Pontificia Universidad
    Cat\'olica de Chile, Casilla 306, Santiago, Chile 6904411 }
\date{Received: \today / Revised version: }

\abstract{ 
  The magnetic arrangement in the vicinity of the interface between a
  ferromagnet and an antiferromagnet is investigated, in particular
  its dependence on the exchange couplings and the temperature.
  Applying a Heisenberg model, both sc(001) and fcc(001) lattices are
  considered and solved by a mean field approximation.  Depending on
  the parameter values a variety of different magnetic configurations
  emerge.  Usually the subsystem with the larger ordering temperature
  induces a magnetic order into the other one (magnetic proximity
  effect).  With increasing temperature a reorientation of the
  magnetic sublattices is obtained.  For coupled sc(001) systems both
  FM and AFM films are disturbed from their collinear magnetic order,
  hence exhibit a similar behavior.  This symmetry is absent for
  fcc(001) films which, under certain circumstances, may exhibit two
  different critical temperatures.  Analytical results are derived for
  simple bilayer systems.
}

\maketitle                                           

\section{Introduction} \label{sec:intro} 
Magnetic reordering in the vicinity of an interface has for a long
time attracted the interest of researchers. In fact, when two
magnetically ordered systems are in atomic contact with each other, 
it is quite natural to expect that in the
vicinity of the interface a novel magnetic arrangement, different from
the bulk one, will set in. This phenomenon is usually referred to as the
magnetic proximity effect (MPE). To the best of our knowledge this
effect was first investigated to treat a ferromagnet 
in contact with a paramagnet \cite{Zuc73}. Since then a vast
literature on the subject has been published, of which we mention just
a few examples \cite{KiZ73}. 

The interest in the MPE has revived lately in relation to the
exchange bias effect \cite{NoS99}.  It occurs when a thin
ferromagnetic (FM) film is deposited on an antiferromagnetic (AFM)
material, resulting in a shift of the hysteresis loop from its normal
(symmetric) position.  If the AFM has a compensated interface
('in-plane AFM'), i.e., if the number of bonds between parallel and 
antiparallel spin pairs across the interface is the same, the AFM
often assumes an almost orthogonal magnetization with respect to
the FM magnetic direction, while the spins of the AFM interface layer
adopt a canted configuration.  This magnetic arrangement of the AFM is
usually called spin-flop-phase, in analogy to an 
AFM system in an external magnetic field \cite{Nee67}.  
The occurrence of the exchange bias effect is, in most likelihood,
related to a certain amount of interface disorder \cite{MNL98}.  

When considering coupled FM-AFM systems we realized that results for 
fully ordered structures are scarce. In previous studies the FM film 
is usually treated as a system with uniform layers, i.e., the spins 
within a given FM layer remain strictly parallel to each other 
\cite{Koo97,HiM86,CrC01,JeD02}. Whereas different magnetization 
directions for different FM layers are considered, each layer rotates 
solidly. We stress that in the case of a compensated FM-AFM interface 
an MPE may be present also for the FM layers close to the interface. 
Thus the magnetic structure of each FM layer is represented, in 
perfect analogy to the AFM layers, by two interpenetrating sublattices 
with different magnetization directions. The consideration of a 
nonuniform intralayer magnetic structure in the FM subsystem leads to 
new features, which in turn are strongly dependent on the underlying 
lattice symmetry. 

\begin{figure*}[t] 
\includegraphics[width=6.5cm]{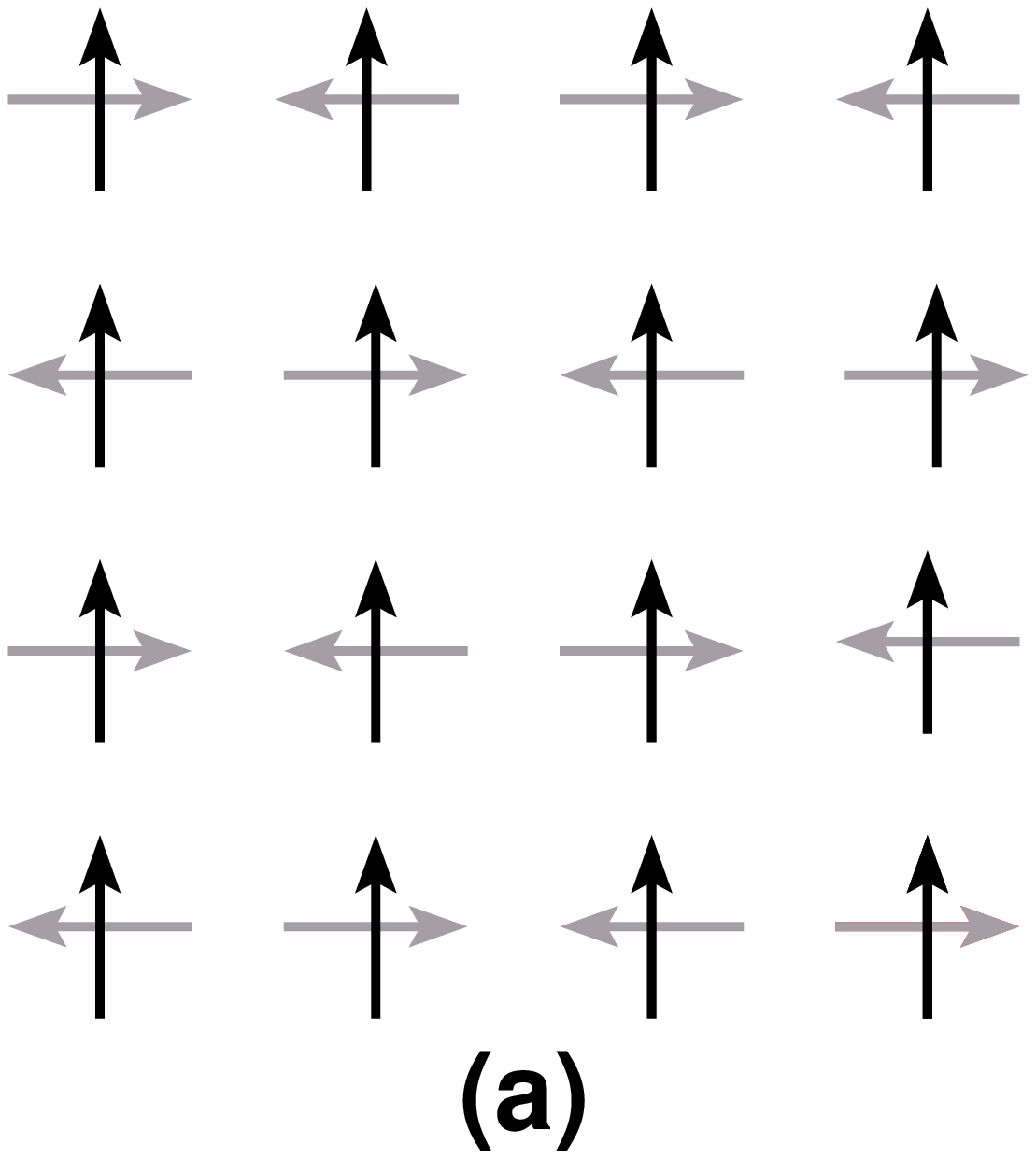} \hspace{1cm} 
\includegraphics[width=9cm]{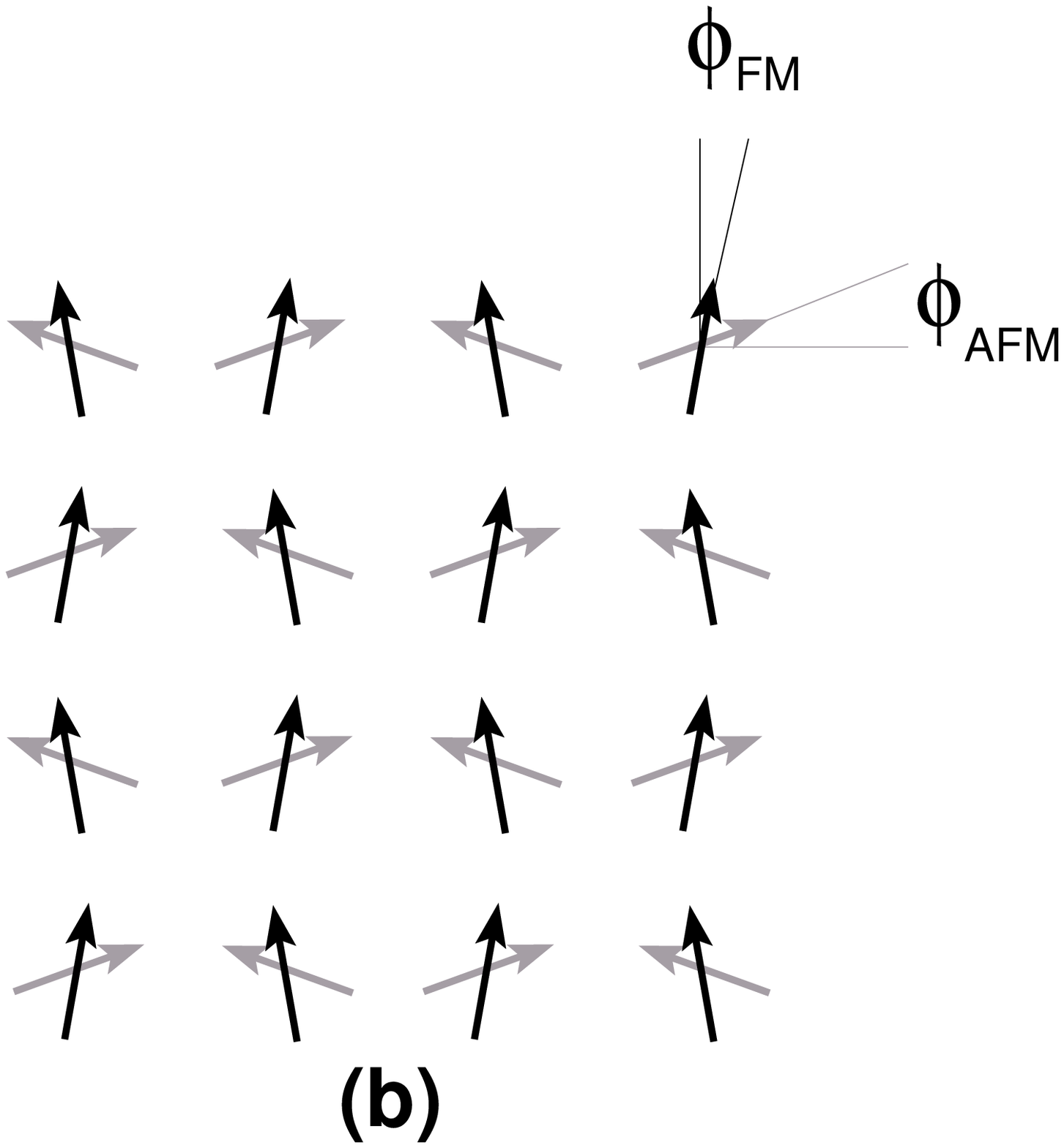} \\[0.5cm]
\caption{  \label{fig1} 
Topview-sketch of the (a) decoupled ($J_\mathrm{int}=0$) 
and (b) coupled ($J_\mathrm{int}>0$) sc(001) bilayer system. 
A single FM layer (dark arrows) and a single AFM layer (grey arrows) 
is assumed, with two sublattices per layer. The angles 
$\phi_\mathrm{FM}$ and $\phi_\mathrm{AFM}$ quantify the deviations 
from the undisturbed magnetic arrangement shown in (a). }
\end{figure*}  

Results concerning the spin reorientation of full magnetic layers have 
been obtained previously for various magnetic systems but, to the best 
of our knowledge, caused by magnetic anisotropies 
\cite{HiM86,CrC01,JeD02}. It is important to stress that although 
we also incorporate anisotropy, the spin rotation in ordered FM-AFM films is 
mainly caused by the \textit{isotropic} exchange interactions. 
Moreover, these systems exhibit a rotation of the \textit{magnetic 
sublattices}, and not a net spin reorientation of the full layers. These 
properties constitute an essential difference of the present treatment 
with respect to previous studies. 

In order to derive a number of general results, while keeping the 
analysis as straightforward as possible, we examine at first the 
magnetic arrangement of a perfectly ordered bilayer consisting of a 
single FM layer that is coupled to a single AFM layer. Using a mean 
field approximation, this particular structure 
yields results which can be written in an \textit{analytical form}. 
In addition, we present a number of results for more realistic 
systems having thicker FM and AFM films. In particular, we investigate 
the effect of the interface coupling on 
the characteristics and magnitude of the MPE at zero and finite 
temperatures. Of special concern is whether, and to which degree, 
magnetic order is induced by the subsystem with the higher (bare) 
ordering (N\'eel or Curie) temperature into the one with the lower 
ordering temperature.  The resulting magnetic arrangements for various 
cases of the bilayer system, for films with several atomic layers, and 
for the corresponding ordering temperatures are determined. In fact, we 
show that, depending on the lattice structure, the proximity effect is 
not always present, and that under certain circumstances two different 
critical temperatures can occur. 

This paper is organized as follows. In Section~\ref{sec:theory} we define 
our physical model. In Section~\ref{sec:zeroT} the magnetic properties 
of the bilayer system at zero temperature are discussed, which
exhibits already a number of general features. Results obtained for 
finite temperatures are presented in Section~\ref{sec:Tne0}. Thicker
films with several FM and AFM layers are considered in 
Section~\ref{sec:thick}. Conclusions are drawn in the last Section. 

\section{Theory} \label{sec:theory} 
To model the magnetic arrangement and ordering temperatures of a 
coupled FM-AFM system we use an XYZ- Heisenberg Hamiltonian 
with localized quantum spins $\mathbf{S}_i$ and spin number $S$, 
\begin{equation} \label{e1} 
\mathcal{H}=-\frac{1}{2}\sum_{\langle i,j\rangle}\,\Big(
J_{ij}\,\mathbf{S}_i \cdot \mathbf{S}_j + 
D_{ij}^x\,S_i^x \cdot S_j^x + D_{ij}^y\,S_i^y \cdot S_j^y \Big) \;. 
\end{equation}
We take into account the isotropic exchange interaction $J_{ij}$ 
between spins located on nearest-neighbor lattice sites $i$ and $j$. 
In addition in-plane easy-axis exchange anisotropies $D_{ij}^x$ and 
$D_{ij}^y$ are considered, which for a particular layer are directed 
either along the $x$- or along the $y$-direction. Note that for 
two-dimensional (2D) magnets a long-range magnetic order at finite 
temperatures exists only in presence of such anisotropies \cite{MeW66}. 
A perfectly ordered layered structure in the $xy$-plane is assumed, 
consisting of an FM film with $n_\mathrm{FM}$ layers and an AFM film 
with $n_\mathrm{AFM}$ layers. Each layer is represented by two 
interpenetrating sublattices, applying otherwise periodic lateral 
boundary conditions. The lattice symmetry, which is assumed to be the 
same for both FM and AFM films, is characterized by the numbers of 
nearest neighbors $z_0$ and $z_1$ within a layer and between adjacent 
layers, respectively. The latter value also refers to the number of 
bonds with which an FM spin is coupled across the interface to 
neighboring spins in the AFM layer. In this study the sc(001) and 
fcc(001) lattices are taken as representative and extremal examples 
corresponding to $z_1=1$ and $z_1=4$, respectively, and $z_0=4$ for 
both symmetries \cite{hexagonal}. As will become apparent in the next 
Sections, the magnetic properties of these two types of coupled FM-AFM 
films differ markedly. 

The FM and AFM subsystems are characterized by the exchange couplings
$J_\mathrm{FM}>0$ and $J_\mathrm{AFM}<0$, and by the usually much 
weaker exchange anisotropies $D_\mathrm{FM}>0$ along the $x$-axis 
and $D_\mathrm{AFM}<0$ along the $y$-axis. Due to shape anisotropy 
the magnetizations of both subsystems are confined to the film plane, 
besides this demagnetizing effect the magnetic dipole interaction is 
not considered explicitely \cite{dipol}. Furthermore, the FM and 
AFM films are coupled across the interface by the interlayer exchange
coupling $J_\mathrm{int}$, where we consider $J_\mathrm{int}>0$ 
without loss of generality, and $D_\mathrm{int}=0$.  
The (unperturbed) ground state for a small interface coupling
$J_\mathrm{int} \to 0$ is defined by a mutually perpendicular
arrangement of the FM and AFM magnetic directions. The choice of the 
anisotropy easy axes support this perpendicular magnetic arrangement.
A net magnetic binding results only if the spins of at least one 
of the subsystems are allowed to deviate from the unperturbed state. 
Hence, the magnetic moments cannot be represented by Ising-like spins. 

In this study we apply a single-spin mean field approximation (MFA). 
Within this method the site-dependent magnetizations 
$\langle\mathbf{S}_i\rangle=\mathbf{M}_i(T)$ with components 
$M_i^x(T)$ and $M_i^y(T)$ are calculated, yielding the magnitudes 
$M_i(T)=|\mathbf{M}_i(T)|$ and in-plane angles 
$\tan\phi_i(T)=\pm M_i^y(T)/M_i^x(T)$. Furthermore, the ordering 
temperatures are determined. For decoupled monolayers 
($J_\mathrm{int}=0$, $n_\mathrm{FM}=n_\mathrm{AFM}=1$) the bare 
Curie temperature $T_C^0$ of the FM and the analogous N\'eel 
temperature $T_N^0$ of the AFM are given by 
\begin{eqnarray} \label{e2a} 
T_C^0 &\;=\;& \frac{\displaystyle S(S+1)}{\displaystyle 3}\;z_0\,
(J_\mathrm{FM}+D_\mathrm{FM}) \;, \nonumber \\[0.3cm] 
\label{e2b} T_N^0 &\;=\;& \frac{\displaystyle S(S+1)}
{\displaystyle 3}\;z_0\,|J_\mathrm{AFM}+D_\mathrm{AFM}| \;, 
\end{eqnarray} 
where the Boltzmann constant $k_B$ is set equal to unity. For thicker 
films the ordering temperature is determined by the largest eigenvalue 
of a particular matrix. We will investigate the MPE for a number of 
different cases, i.e., whether and to which degree a magnetic order 
propagates from the subsystem with the larger bare ordering 
temperature into the other one. The corresponding magnetic structure 
is characterized by the magnetization vectors $\mathbf{M}_i(T)$. As 
mentioned, at first we will consider the particularly simple bilayer 
system ($n_\mathrm{FM}=n_\mathrm{AFM}=1$) which allows to draw a 
number of general results and analytical expressions. Later on we take 
into account coupled FM-AFM systems with thicker films. Since the 
explicit consideration of anisotropies is not decisive within MFA, for 
simplicity we include them into the exchange couplings: 
$J_\mathrm{FM}+D_\mathrm{FM}\to J_\mathrm{FM}$ and 
$J_\mathrm{AFM}+D_\mathrm{AFM}\to J_\mathrm{AFM}$. 
For the spin quantum number we use $S=1$ throughout. 

\begin{figure}[t] 
\includegraphics[width=5.22cm,bb=50 25 500 760,angle=-90,clip]{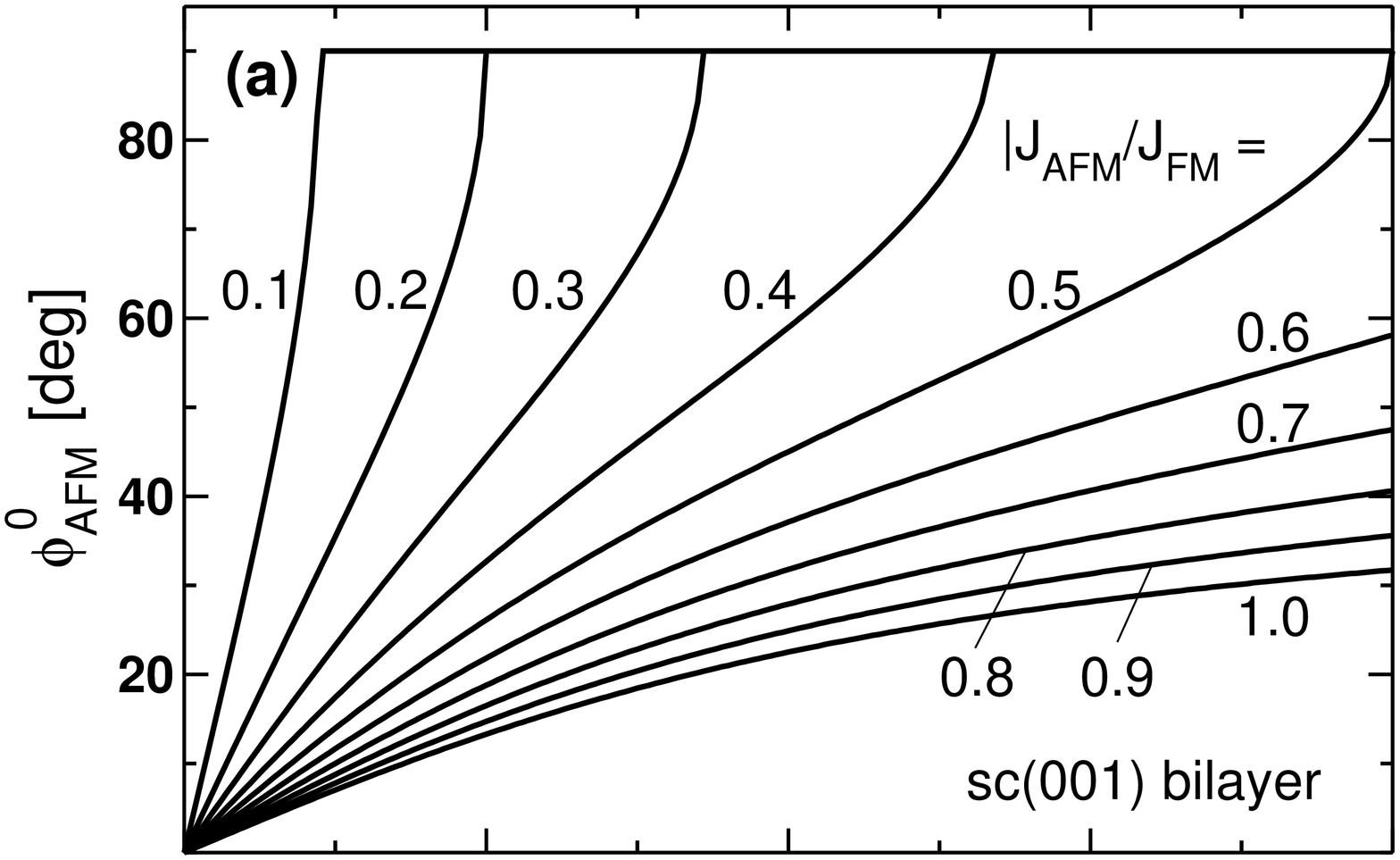} \\
\includegraphics[width=6cm,bb=50 25 565 760,angle=-90,clip]{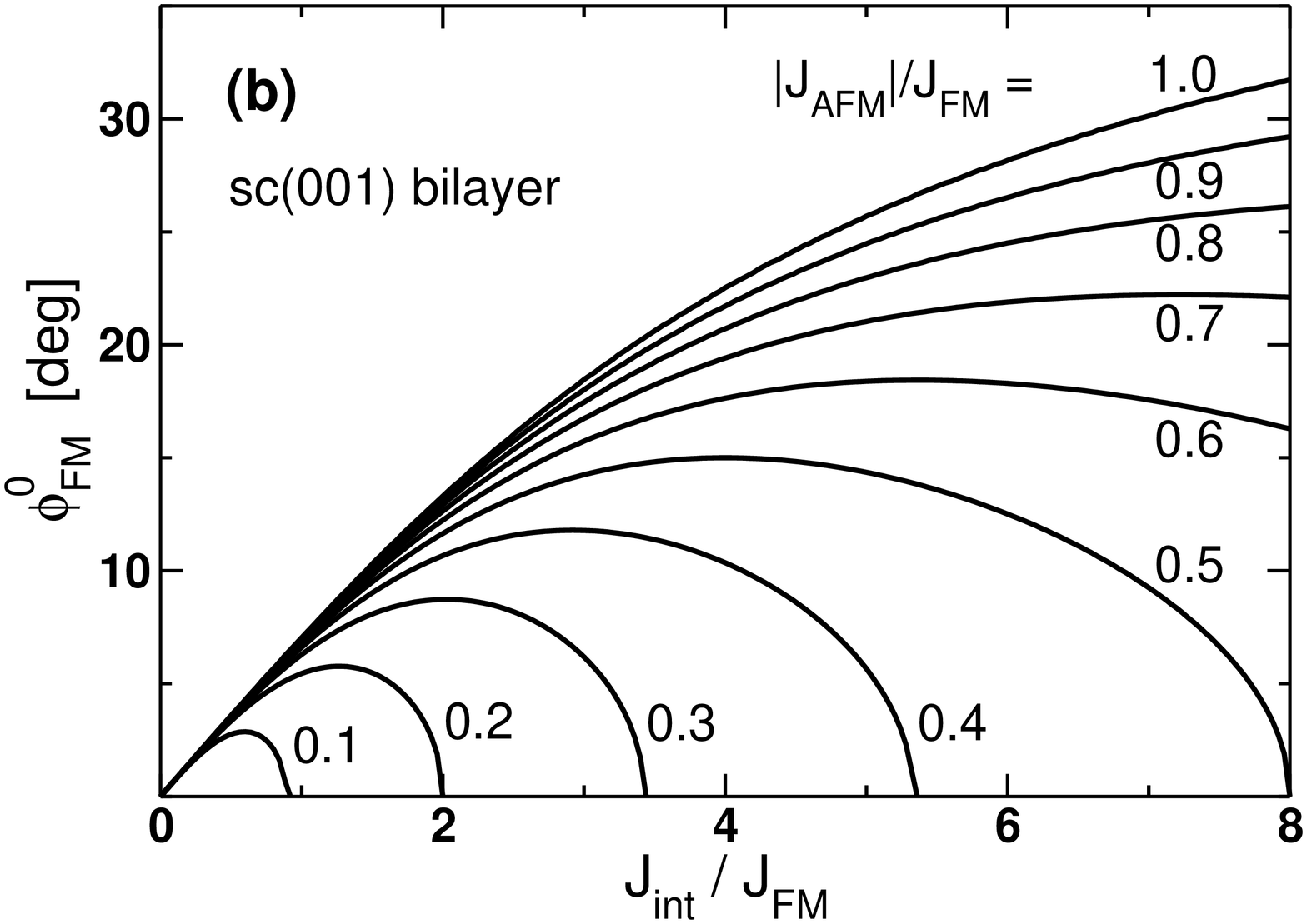} \\ 
\caption{\label{fig2} Equilibrium angles (a) $\phi_\mathrm{AFM}^0$ of the 
  AFM layer and (b) $\phi_\mathrm{FM}^0$ of the FM layer 
  as functions of the interlayer exchange coupling $J_\mathrm{int}$ 
  for an sc(001) bilayer at $T=0$. The different plots correspond to 
  different AFM exchange couplings $J_\mathrm{AFM}$. $J_\mathrm{FM}$ 
  is the exchange in the FM layer, and $z_0$ and $z_1$ the numbers of 
  nearest neighbors within a layer and between adjacent layers. }
\end{figure}

\begin{figure*}[t] 
\includegraphics[width=6.5cm]{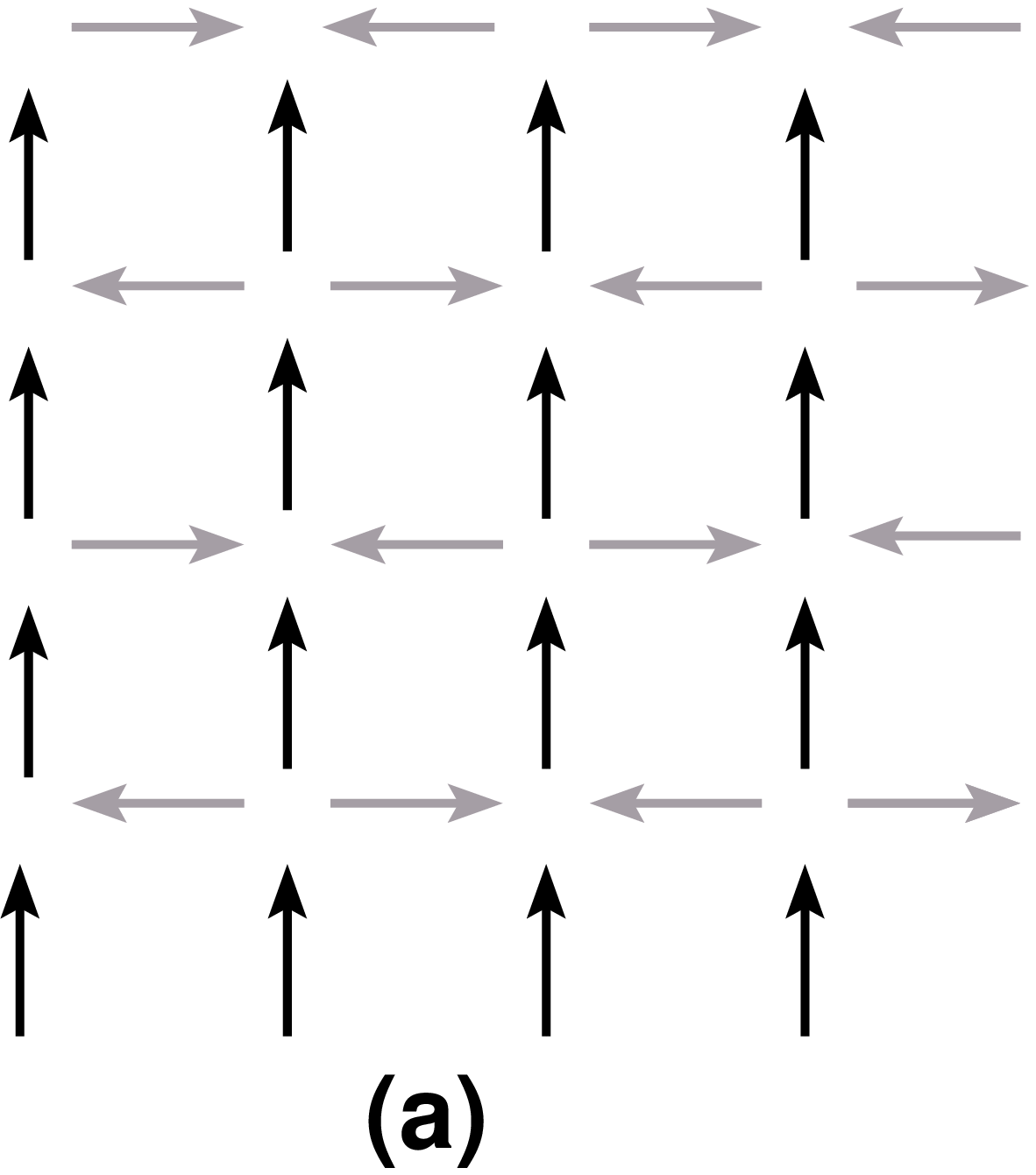} \hspace{1cm} 
\includegraphics[width=9cm]{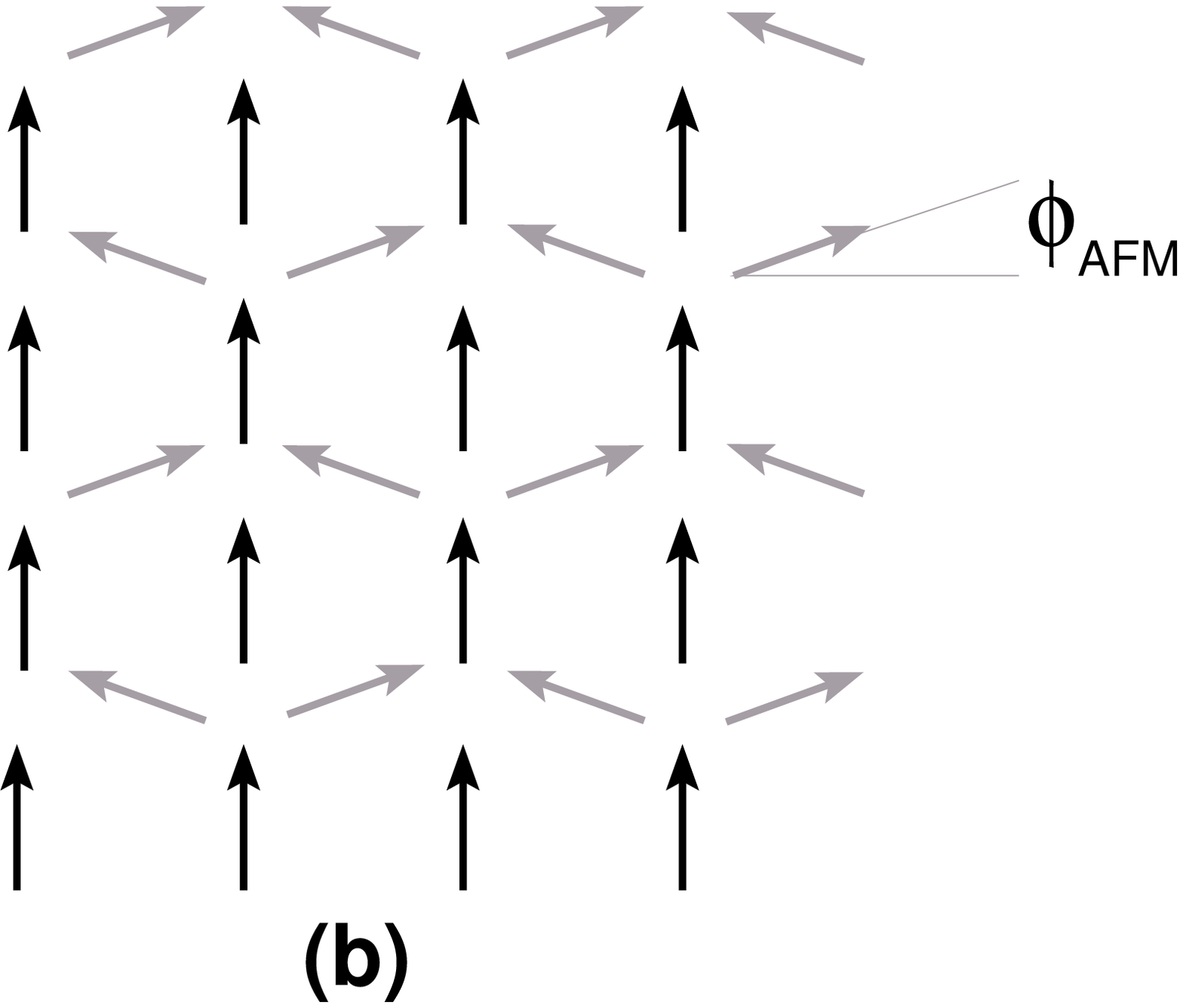} \\[0.5cm]
\caption{ \label{fig3} 
  Same as Figure~\ref{fig1} for an fcc(001) bilayer. The 
  angle $\phi_\mathrm{AFM}$ quantifies the deviation from the 
  undisturbed AFM arrangement. The FM layer remains collinear. } 
\end{figure*} 

\section{Bilayers: Zero temperature} \label{sec:zeroT}
\subsection{sc(001) -- bilayer} \label{sc001bilayer}
For this lattice type \textit{both} the FM and AFM layers are disturbed 
from their ground state, thus also the FM layer `dimerizes' and 
exhibits a noncollinear magnetization. The undisturbed magnetic 
arrangement of an sc(001) bilayer is depicted in Figure~\ref{fig1}a. 
For $J_\mathrm{int}>0$ both FM and AFM layers assume 
a canted magnetic arrangement, as sketched in Figure~\ref{fig1}b. 
The canting angles $\phi_\mathrm{FM}$ and $\phi_\mathrm{AFM}$ 
represent the deviations from the decoupled bilayer. 

The energy of such an arrangement is given by
\begin{eqnarray} \label{e2} 
E_\mathrm{sc(001)}(\phi_\mathrm{FM},\phi_\mathrm{AFM}) &=& 
\nonumber \\ && \hspace*{-3.5cm}
-\frac{z_0}{2}\,J_\mathrm{FM}\,\cos(2\,\phi_\mathrm{FM})
-\frac{z_0}{2}\,|J_\mathrm{AFM}|\, \cos(2\,\phi_\mathrm{AFM}) 
\nonumber \\ && \hspace*{-3.5cm} -z_1\,J_\mathrm{int}\,
\cos(\pi/2-\phi_\mathrm{FM}-\phi_\mathrm{AFM}) \;. 
\end{eqnarray}
Differentiation of 
$E_\mathrm{sc(001)}(\phi_\mathrm{FM},\phi_\mathrm{AFM})$ with respect
to $\phi_\mathrm{FM}$ and $\phi_\mathrm{AFM}$ yields the conditions for
the equilibrium angles $\phi_\mathrm{FM}^0$ and $\phi_\mathrm{AFM}^0$, 
\begin{eqnarray} \label{e4} 
z_0\,J_\mathrm{FM}\,\sin(2\,\phi_\mathrm{FM}^0) &=& 
z_0\,|J_\mathrm{AFM}|\,\sin(2\,\phi_\mathrm{AFM}^0) 
\nonumber \\ && \hspace*{-2cm} 
=z_1\,J_\mathrm{int}\,\cos(\phi_\mathrm{FM}^0+
\phi_\mathrm{AFM}^0) \;. 
\end{eqnarray}
We emphasize that this behavior refers to a magnetic rotation of the
two sublattices, with angles $\phi_i^0$ and $\pi-\phi_i^0$, and not to
a net spin reorientation of layer $i$. 

First we consider $|J_\mathrm{AFM}|<J_\mathrm{FM}$. In
Figure~\ref{fig2} the angles $\phi_\mathrm{FM}^0$ and 
$\phi_\mathrm{AFM}^0$ are shown as functions of the interlayer
coupling $J_\mathrm{int}$ for different values of $|J_\mathrm{AFM}|$. 
The following properties are quite apparent: 
\begin{itemize}
\item For a small $|J_\mathrm{AFM}|$ the AFM spins 
  quickly turn into the direction of the FM as
  $J_\mathrm{int}$ increases. A parallel orientation of the AFM spins 
  with respect to the FM, i.e., $\phi_\mathrm{AFM}^0=90^\circ$ and 
  $\phi_\mathrm{FM}^0=0^\circ$, is reached at the particular strength 
  $J_\mathrm{int}^\|$ of the interlayer coupling, given by 
\begin{equation}
J_\mathrm{int}^\|=\frac{z_0}{z_1}\;
\frac{2\,J_\mathrm{FM}\,|J_\mathrm{AFM}|}{J_\mathrm{FM}-|J_\mathrm{AFM}|} 
\;. \label{e6}  \end{equation}
  The larger $|J_\mathrm{AFM}|$ the larger is the value of
  $J_\mathrm{int}^\|$ required to reach that limit.
\item With increasing $J_\mathrm{int}$ the FM angle $\phi_\mathrm{FM}^0$ 
  increases and exhibits a maximum at 
\begin{equation} 
J_\mathrm{int}^\mathrm{max}=\frac{z_0}{z_1}\;|J_\mathrm{AFM}|\;
\sqrt{\frac{2\,J_\mathrm{FM}}{J_\mathrm{FM}-|J_\mathrm{AFM}|}} 
\;, \label{e7}  \end{equation}
  assuming the value 
  $\sin(2\,\phi_\mathrm{FM}^\mathrm{max})=|J_\mathrm{AFM}|/J_\mathrm{FM}$  
  and coinciding with $\phi_\mathrm{AFM}^0=45^\circ$. Notice that 
  $\phi_\mathrm{FM}^0(J_\mathrm{int})$ and 
  $\phi_\mathrm{AFM}^0(J_\mathrm{int})$ in general are not symmetric 
  with respect to $J_\mathrm{int}^\mathrm{max}$. 
\item For the limiting case $|J_\mathrm{AFM}|=J_\mathrm{FM}$, 
  no maximum of $\phi_\mathrm{FM}^0$ is obtained.
  Instead one has $\tan(2\,\phi_\mathrm{FM}^0)=
  \tan(2\,\phi_\mathrm{AFM}^0)=(z_1\,J_\mathrm{int})/(z_0\,J_\mathrm{FM})$ 
  For $J_\mathrm{int}\to\infty$ one obtains 
  $\phi_\mathrm{FM}^0=\phi_\mathrm{AFM}^0=45^\circ$.  
\item For $J_\mathrm{int}<0$ the same results emerge, if one
  performs the transformations 
  $\phi_\mathrm{FM}^0\to-\phi_\mathrm{FM}^0$ and
  $\phi_\mathrm{AFM}^0\to-\phi_\mathrm{AFM}^0$. 
\item The sc(001) bilayer is characterized by an apparent symmetry
  between the FM and AFM layers as determined within MFA.  For
  $J_\mathrm{FM}<|J_\mathrm{AFM}|$ the 
  behavior of the FM and AFM layers, in particular the equilibrium
  angles $\phi_\mathrm{FM}^0$ and $\phi_\mathrm{AFM}^0$, becomes
  interchanged, as can be seen from the symmetry of equations~(\ref{e2})
  and (\ref{e4}). If one exchanges $J_\mathrm{FM}$ and 
  $|J_\mathrm{AFM}|$ in the preceding deduction, Figure~\ref{fig2}a 
  is valid for $\phi_\mathrm{FM}^0$ and Figure~\ref{fig2}b for 
  $\phi_\mathrm{AFM}^0$. Thus, for $J_\mathrm{int}>J_\mathrm{int}^\|$, 
  antiferromagnetic order of the FM layer on top of the undisturbed
  AFM layer sets
  in. Another system exhibiting this behavior is the bcc(110) bilayer. 
\end{itemize} 

\begin{figure}[h] 
\includegraphics[width=6cm,bb=50 25 575 760,angle=-90,clip]{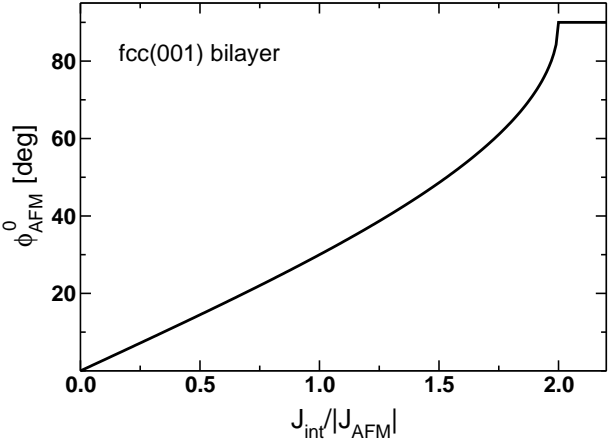} \\
\caption{\label{fig4} Equilibrium angle $\phi_\mathrm{AFM}^0$ of the 
  AFM layer as function of the interlayer exchange $J_\mathrm{int}$ for
  an fcc(001) bilayer at $T=0$.  The corresponding angle of
  the FM layer is $\phi_\mathrm{FM}^0=0$. } 
\end{figure}

\subsection{fcc(001) -- bilayer}  
The fcc(001) bilayer is characterized by the fact that for an
undisturbed AFM the sum of the coupling energies to a given FM spin 
vanishes, as can be seen from the undisturbed arrangement illustrated 
in Figure~\ref{fig3}a. By setting up an equation similar to 
equation~(\ref{e2}) one can show that $\phi_\mathrm{FM}^0=0$, hence in 
this case the spin structure of the FM always remains strictly collinear.
The resulting magnetic structure of a coupled fcc(001) bilayer is
shown in Figure~\ref{fig3}b. Thus, the symmetry between the FM
and AFM subsystems of the sc(001) bilayer is no longer present for the
fcc(001) one. This is a consequence of the fact that for the sc(001) 
bilayer each FM spin couples to a \textit{single} AFM sublattice,
while for the fcc(001) interface each FM spin couples identically to 
\textit{both} AFM sublattices. A similar behavior holds for bcc(001) 
films. 

The corresponding energy expression $E_\mathrm{fcc(001)}$ is 
obtained from equation~(\ref{e2}) by setting $\phi_\mathrm{FM}=0$.  
Differentiation with respect to $\phi_\mathrm{AFM}$ yields 
the equilibrium angle $\phi_\mathrm{AFM}^0$ of the
disturbed AFM spin arrangement, 
\begin{equation} \label{e9} 
\sin(\phi_\mathrm{AFM}^0)=\frac{z_1\,J_\mathrm{int}}
{2\,z_0\,|J_\mathrm{AFM}|} \;, 
\end{equation}
which is shown in Figure~\ref{fig4} as function of $J_\mathrm{int}$. For 
$z_1\,J_\mathrm{int}>2\,z_0\,|J_\mathrm{AFM}|$ one obtains 
$\phi_\mathrm{AFM}^0=90^\circ$, i.e., the spins of the AFM layer order 
parallel to the FM ones. The case $J_\mathrm{int}<0$ is recovered by 
replacing $\phi_\mathrm{AFM}^0\to-\phi_\mathrm{AFM}^0$.  

\begin{figure}[t] 
\includegraphics[width=5cm,bb=45 25 500 760,angle=-90,clip]{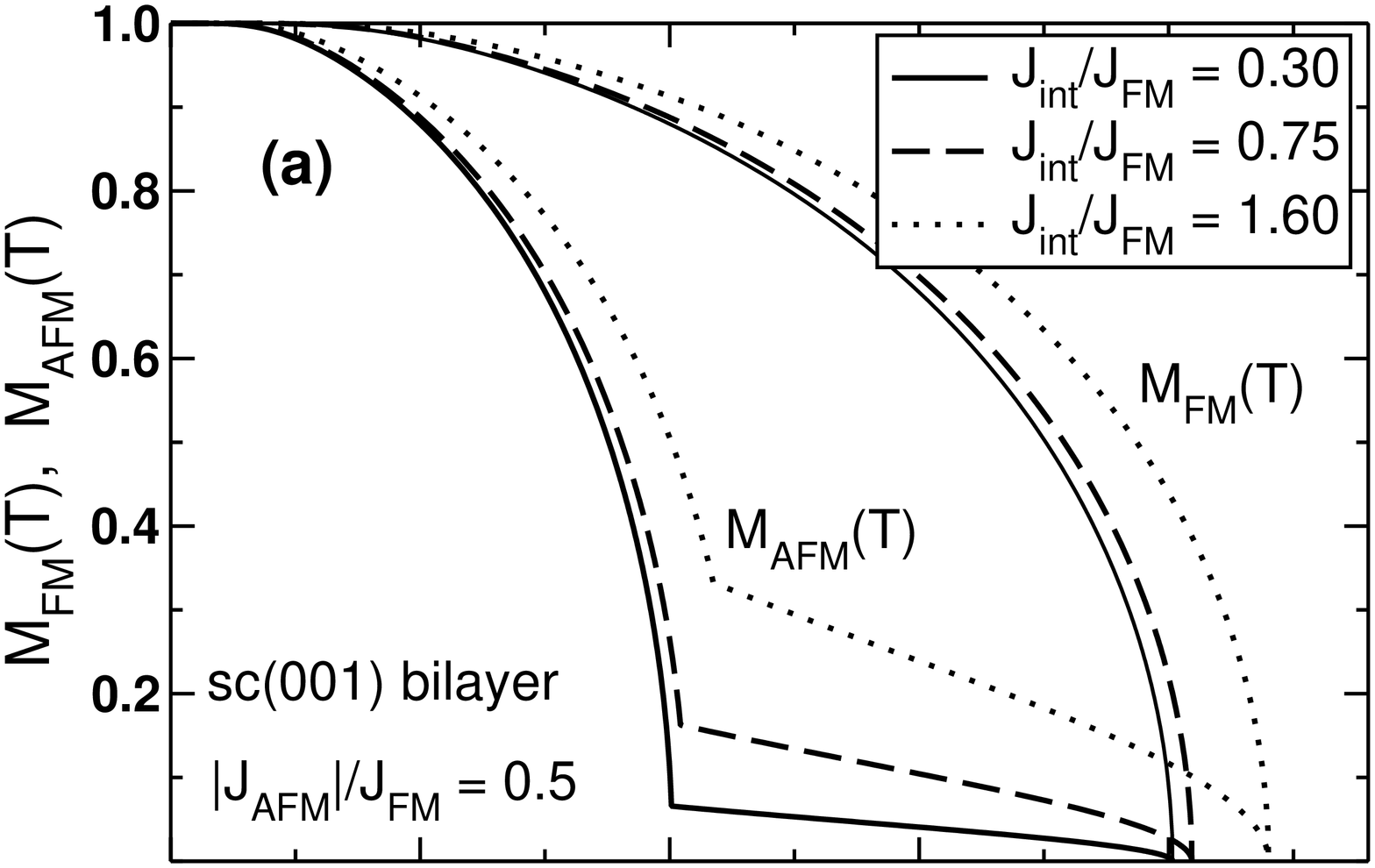} \\
\includegraphics[width=6cm,bb=45 25 590 760,angle=-90,clip]{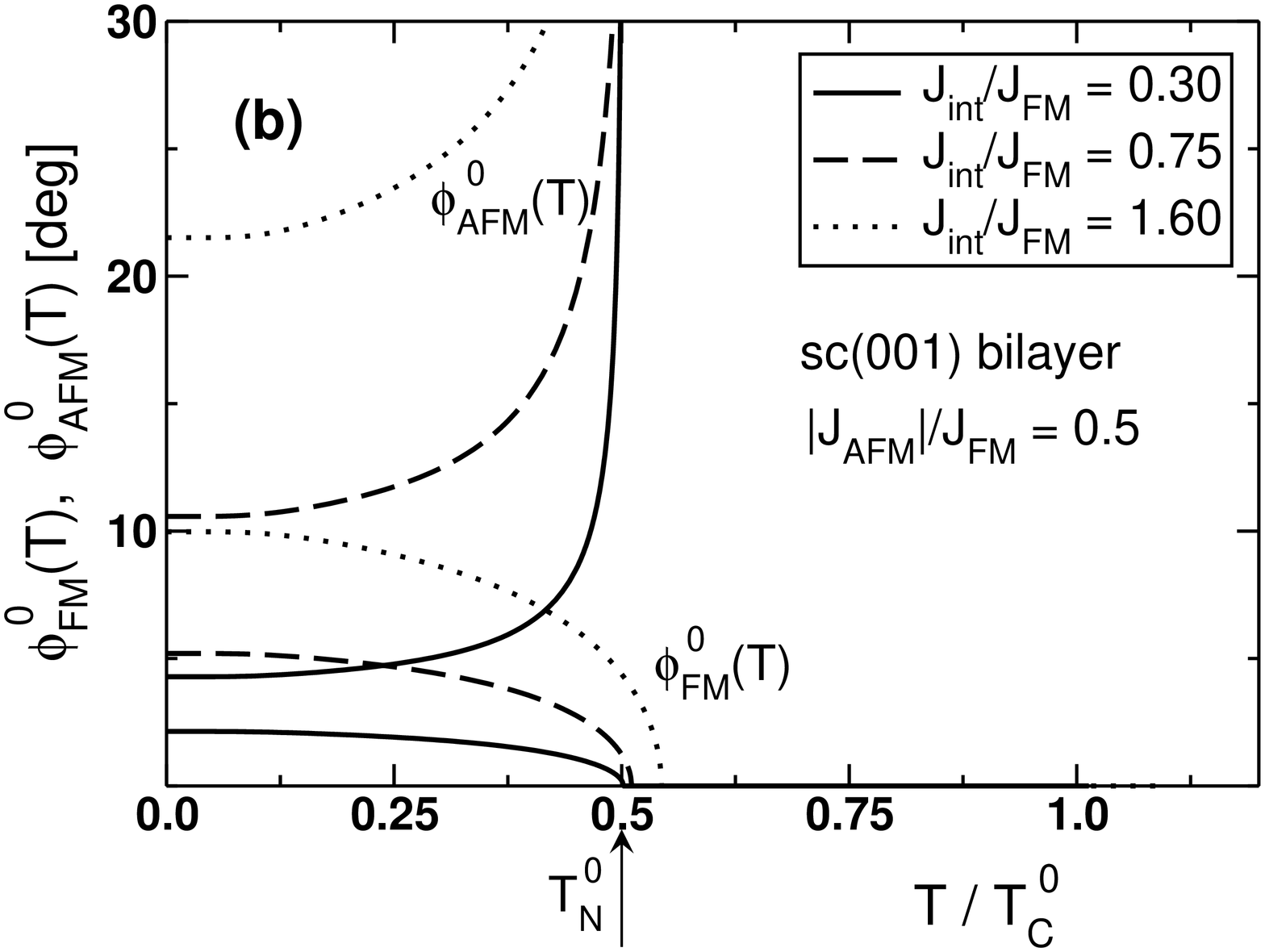} \\
\caption{\label{fig5} (a) Magnetizations 
  $M_i(T)$ and (b) equilibrium angles 
  $\phi_i^0(T)$ for an sc(001) bilayer as functions of the temperature 
  $T$ for different values of the interlayer exchange coupling
  $J_\mathrm{int}$. The AFM exchange is chosen to be
  $|J_\mathrm{AFM}|/J_\mathrm{FM}=0.5$, hence $T_N^0 < T_C^0$. The 
  temperature is given in units of the bare Curie temperature $T_C^0$
  of the FM monolayer. At the sublattice reorientation temperature 
  $T_R(J_\mathrm{int})\geq T_N^0$ one obtains 
  $\phi_\mathrm{AFM}^0(T)\to90^\circ$ and $\phi_\mathrm{FM}^0(T)\to 0$.}
\end{figure} 

\section{Bilayer: Finite temperatures} \label{sec:Tne0} 
We now turn our attention to the magnetic arrangement of the coupled
FM-AFM bilayer at finite temperatures. Like in the previous 
Section we distinguish between an sc(001) and an fcc(001) symmetry. 
Furthermore, we treat the cases $T_N^0<T_C^0$, $T_N^0>T_C^0$, and 
$T_N^0=T_C^0$ separately.

Let us at first present the ordering temperature $T_C$ for a coupled 
magnetic bilayer with a collinear magnetization. Its two layers are 
characterized by the exchange couplings $J_1$ and $J_2$, which can be 
of either sign. Within MFA one obtains 
\begin{eqnarray} \label{e14}
T_C &=& \frac{S(S+1)}{6}\bigg[z_0(\,J_1+J_2) \nonumber \\
&& +\sqrt{z_0^2\,(J_1-J_2)^2+4\,(z_1\,J_\mathrm{int})^2}\bigg] \;. 
\end{eqnarray}
Except for the cases that will be mentioned below, $T_C$ 
of the coupled bilayer is always larger than the largest 
bare ordering temperature ($T_N^0$ or $T_C^0$) of the decoupled 
monolayers, regardless of the relative magnitude of $J_1$ and $J_2$, 
and of the sign of $J_\mathrm{int}$. For unequal layers 
($J_1\neq J_2$) and a small coupling $J_\mathrm{int}$ one obtains an 
increase of $T_C$ given approximately by 
\begin{equation} \label{e13} 
\Delta T_C(J_\mathrm{int}) \simeq \frac{S(S+1)}{3}\; 
\frac{(z_1\,J_\mathrm{int})^2}{z_0\,|J_1-J_2|} \;. 
\end{equation}
From the denominator of equation~(\ref{e13}) one observes that the 
increase of $T_C$ for an FM bilayer ($J_1,J_2>0$) 
will be larger than the one for a corresponding FM-AFM bilayer 
($J_1>0$, $J_2<0$). Within MFA the results for 
$J_\mathrm{int}<0$ are identical to the corresponding ones for 
$J_\mathrm{int}>0$, if the signs of $\phi_\mathrm{FM}^0(T)$ and 
$\phi_\mathrm{AFM}^0(T)$ are adapted appropriately. 

\subsection{sc(001) -- bilayer} \label{subsec:sc001} 
\noindent a) $T_N^0<T_C^0$. For the AFM coupling we choose 
$|J_\mathrm{AFM}|/J_\mathrm{FM}$ $=0.5$. In Figure~\ref{fig5} we display 
the magnetizations $M_\mathrm{FM}(T)$
and $M_\mathrm{AFM}(T)$, and the corresponding equilibrium angles
$\phi_\mathrm{FM}^0(T)$ and $\phi_\mathrm{AFM}^0(T)$, as functions of
the temperature $T$.  Different values of the interlayer
coupling $J_\mathrm{int}>0$ are used as indicated. At low
temperatures both subsystems deviate from the undisturbed magnetic
arrangement. With increasing temperature the equilibrium angle
$\phi_\mathrm{FM}^0(T)$ of the FM layer decreases, whereas
$\phi_\mathrm{AFM}^0(T)$ of the AFM layer increases. 
Approaching the \textit{sublattice reorientation temperature} $T_R$, 
given implicitly by the relation
\begin{eqnarray} \label{e10} 
 && z_1\,J_\mathrm{int}\;
\big[J_\mathrm{FM}\;M_\mathrm{FM}^2(T_R)-|J_\mathrm{AFM}|\;
M_\mathrm{AFM}^2(T_R)\big] \nonumber \\
&=& 2\,z_0\,J_\mathrm{FM}\;|J_\mathrm{AFM}|\;
M_\mathrm{FM}(T_R)\;M_\mathrm{AFM}(T_R) \;, 
\end{eqnarray} 
the AFM spins turn into the direction of the FM spins, and one obtains 
$\phi_\mathrm{FM}^0(T)\to0^\circ$, 
$\phi_\mathrm{AFM}^0(T)\to90^\circ$. Thus, for $T>T_R$ the AFM layer 
adopts ferromagnetic order. $M_\mathrm{AFM}(T)$ exhibits a sharp kink 
at $T_R$, whereas $M_\mathrm{FM}(T)$ shows no particular features. The 
FM-AFM bilayer becomes paramagnetic above the ordering temperature 
$T_C$ given by equation~(\ref{e14}). 

\noindent b) $T_N^0>T_C^0$. Here the strength of the AFM coupling 
is stronger than the FM one. We adopt $|J_\mathrm{AFM}|/J_\mathrm{FM}=2$ 
for comparison with the previous case. Then  
the results for $T_N^0>T_C^0$ are fully symmetric with the ones 
derived for $T_N^0<T_C^0$ shown in Figure~\ref{fig5}, if one 
interchanges $M_\mathrm{FM}(T)\,\leftrightarrow\,M_\mathrm{AFM}(T)$, 
$\phi_\mathrm{FM}^0(T)\,\leftrightarrow\,\phi_\mathrm{AFM}^0(T)$, and 
$T_N^0\,\leftrightarrow\,T_C^0$. Thus new figures are not required. 
Notice that for $T>T_R$ the FM layer assumes an antiferromagnetic 
structure. This finding demonstrates the symmetry of the FM and AFM 
layers of the sc(001) lattice within MFA, which also holds for finite 
temperatures. 

\noindent c) $T_N^0=T_C^0$. For the particular case 
$J_\mathrm{AFM}=-J_\mathrm{FM}$ the angles 
$\phi_\mathrm{FM}^0(T)$ and $\phi_\mathrm{AFM}^0(T)$ are independent
of the temperature and are given by
$\tan(2\,\phi_\mathrm{FM}^0)=\tan(2\,\phi_\mathrm{AFM}^0)=
(z_1\,J_\mathrm{int})/(z_0\,J_\mathrm{FM})$. The magnetizations 
$M_\mathrm{FM}(T)$ and \break $M_\mathrm{AFM}(T)$ are identical and 
vanish at the ordering temperature, cf.\ equation~(\ref{e14}), 
\begin{equation} \label{e14a} 
T_C=\frac{S(S+1)}{3}\;z_0\,J_\mathrm{FM}\; 
\sqrt{1+j_\mathrm{int}^2} \;. 
\end{equation}

\begin{figure}[t] 
\includegraphics[width=5.23cm,bb=45 25 505 760,angle=-90,clip]{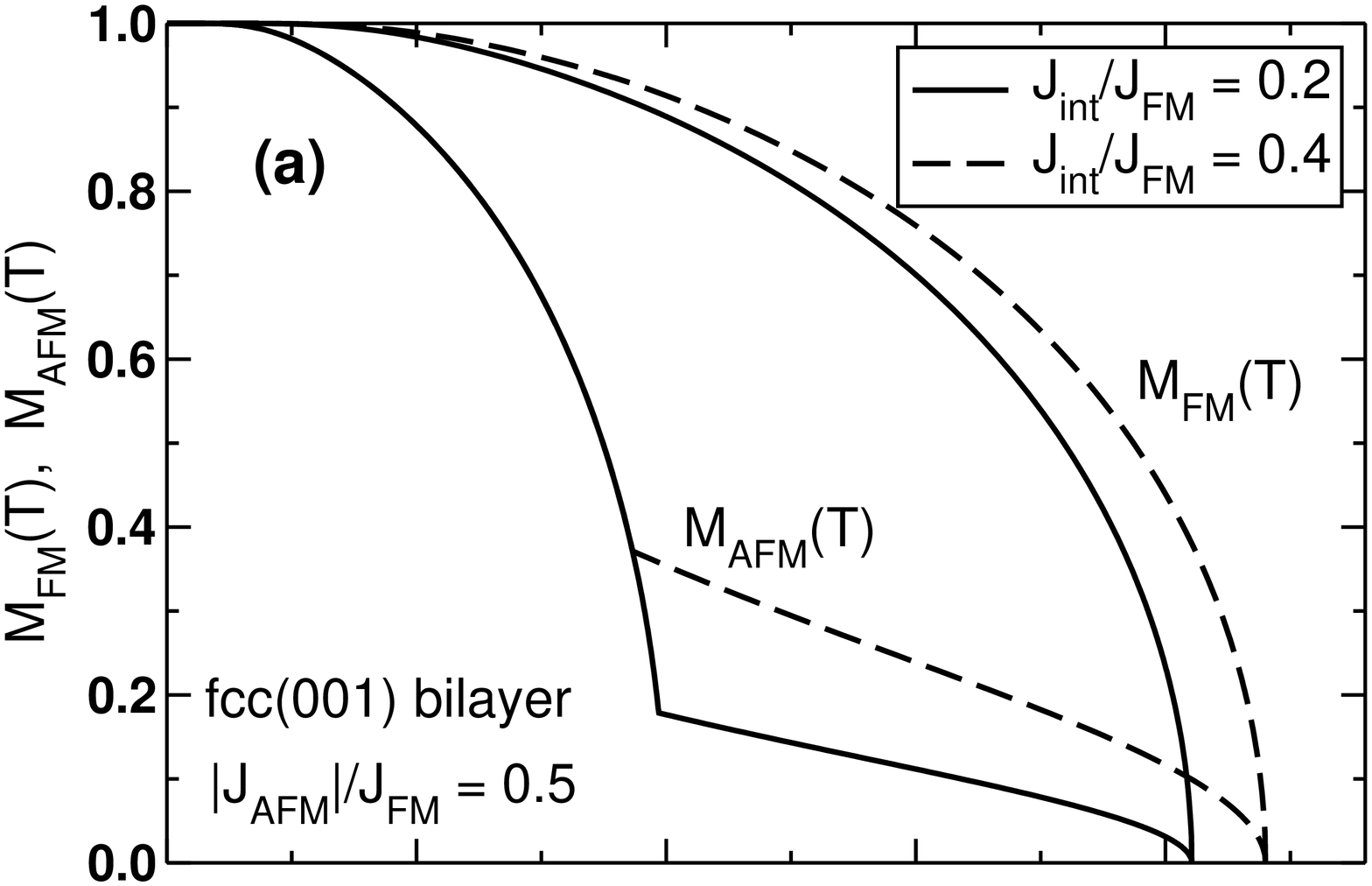} \\
\includegraphics[width=6cm,bb=50 25 575 760,angle=-90,clip]{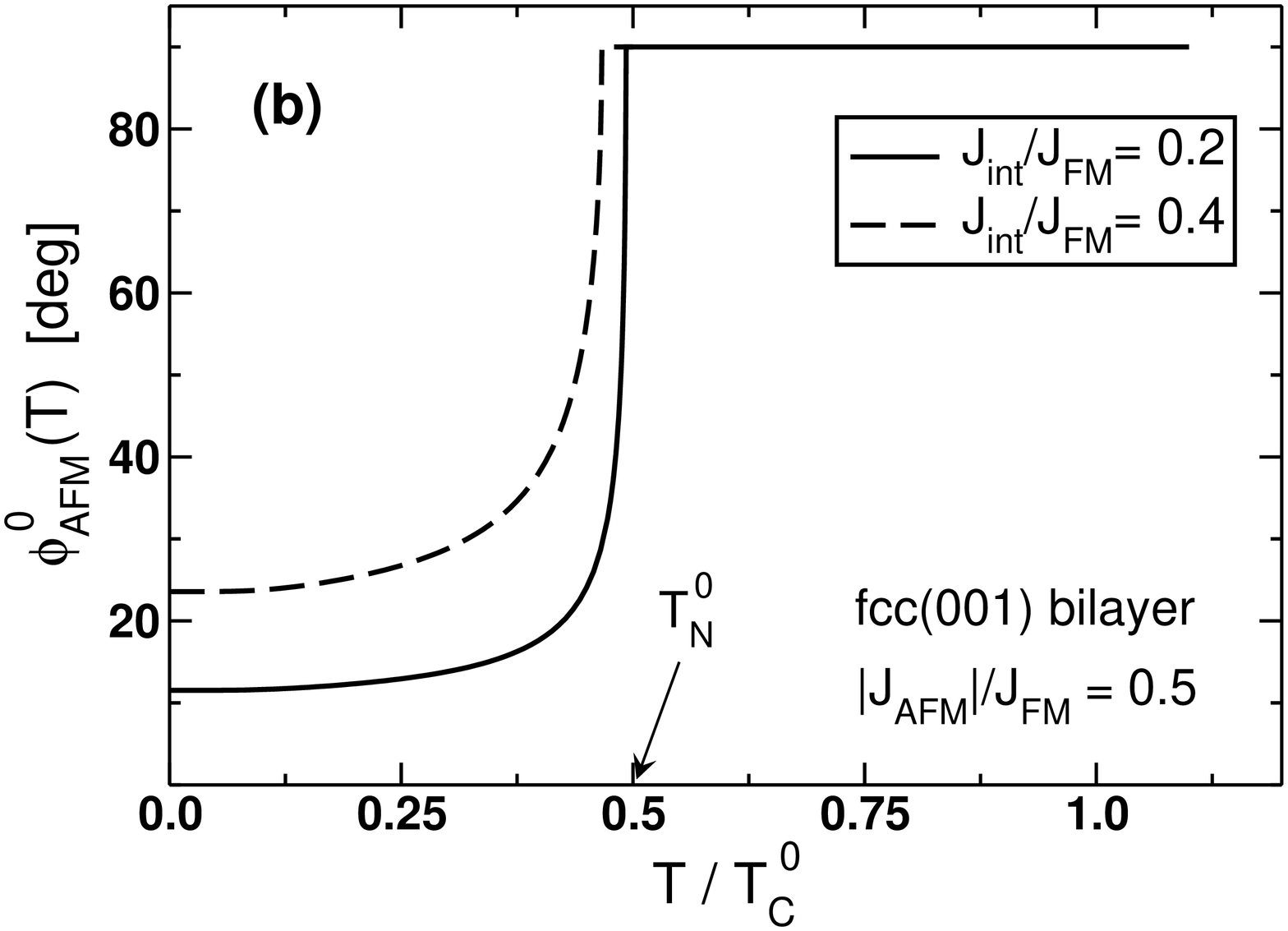} \\
\caption{\label{fig6} (a) Magnetizations $M_i(T)$ and (b) 
  AFM equilibrium angles $\phi_\mathrm{AFM}^0(T)$ of an fcc(001)
  bilayer as functions of the temperature $T/T_C^0$ for different values 
  of $J_\mathrm{int}$. The AFM exchange is put equal to  
  $|J_\mathrm{AFM}|/J_\mathrm{FM}=0.5$, hence $T_N^0<T_C^0$. 
  The FM angle is $\phi_\mathrm{FM}^0(T)=0$.}
\end{figure}

\subsection{fcc(001) -- bilayer}  
As mentioned in Section~\ref{sec:zeroT}, the behavior of the fcc(001) 
bilayer system is not symmetric, which also holds for finite 
temperatures. Several cases have to be distinguished, for this purpose 
we define the \textit{crossover interlayer coupling} 
\begin{equation} \label{e15} 
J_\mathrm{int}^*= \frac{z_0}{z_1}\;
\sqrt{2\,|J_\mathrm{AFM}|\;\big(|J_\mathrm{AFM}|-J_\mathrm{FM}\big)} \;. 
\end{equation}

\noindent a) $T_N^0<T_C^0$. This case is similar to the analogous
sc(001) one. However, unlike that system, for the fcc(001) bilayer the 
spins in the FM layer remain always collinear, i.e., 
$\phi_\mathrm{FM}^0(T)=0$. With increasing temperature the equilibrium 
angle $\phi_\mathrm{AFM}^0(T)$ of the AFM layer increases and 
approaches $90^\circ$ for the temperature $T_R$ given by 
\begin{equation} \label{e10a} 
z_1\,J_\mathrm{int}\;M_\mathrm{FM}(T_R)=2\,z_0\,|J_\mathrm{AFM}|\; 
M_\mathrm{AFM}(T_R) \;. 
\end{equation}  
For $T>T_R$ the AFM spins 
remain in a ferromagnetic structure up to the ordering 
temperature given by equation~(\ref{e14}). This behavior is 
depicted in Figure~\ref{fig6} for different values of $J_\mathrm{int}$. 
Notice that due to the larger number $z_1$ of interlayer bonds 
the influence of the interlayer coupling for the fcc(001) bilayer 
is more pronounced as compared to the sc(001) system. 

\begin{figure}[t] 
\includegraphics[width=5.22cm,bb=45 25 500 760,angle=-90,clip]{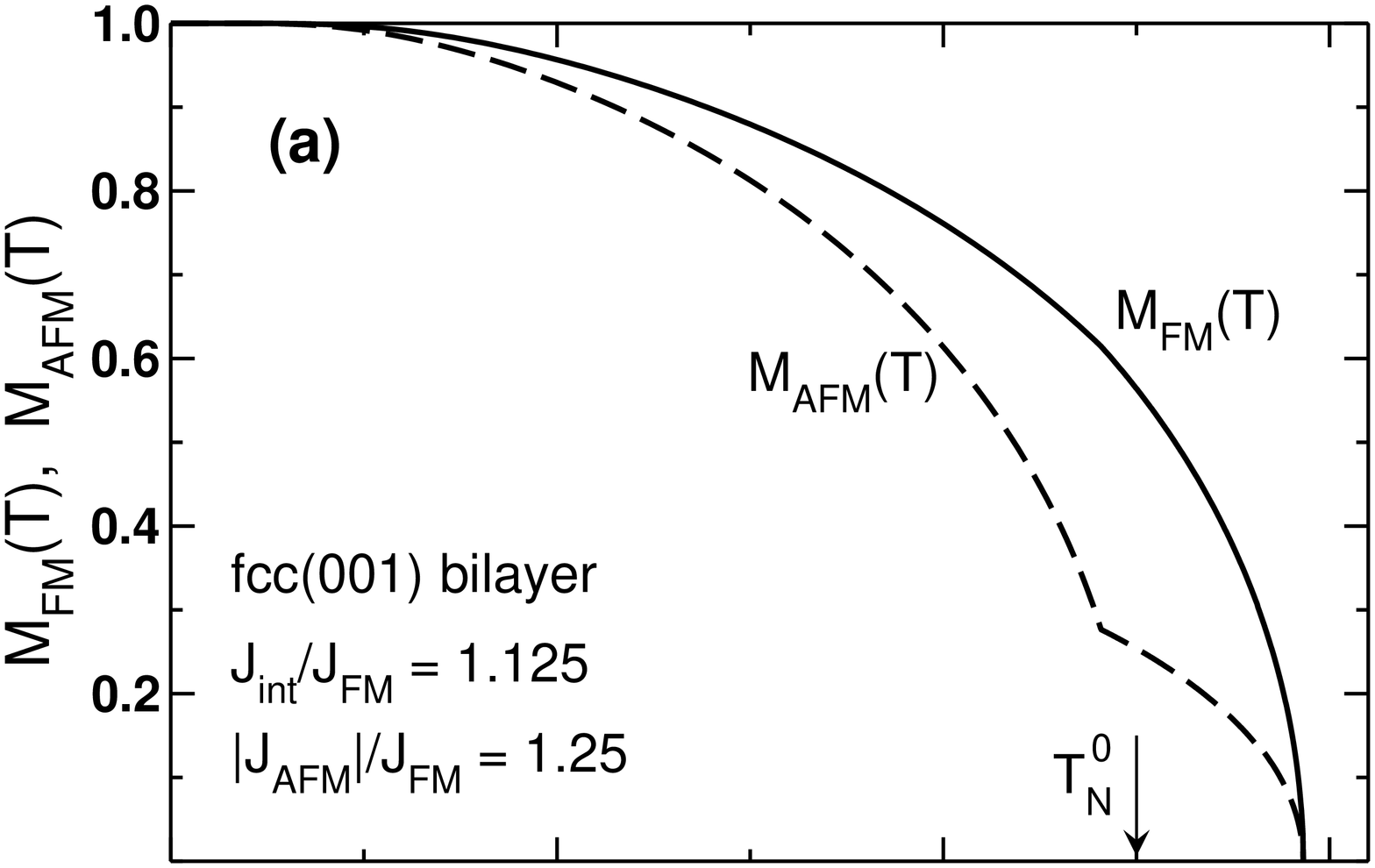} \\
\includegraphics[width=6cm,bb=50 25 575 760,angle=-90,clip]{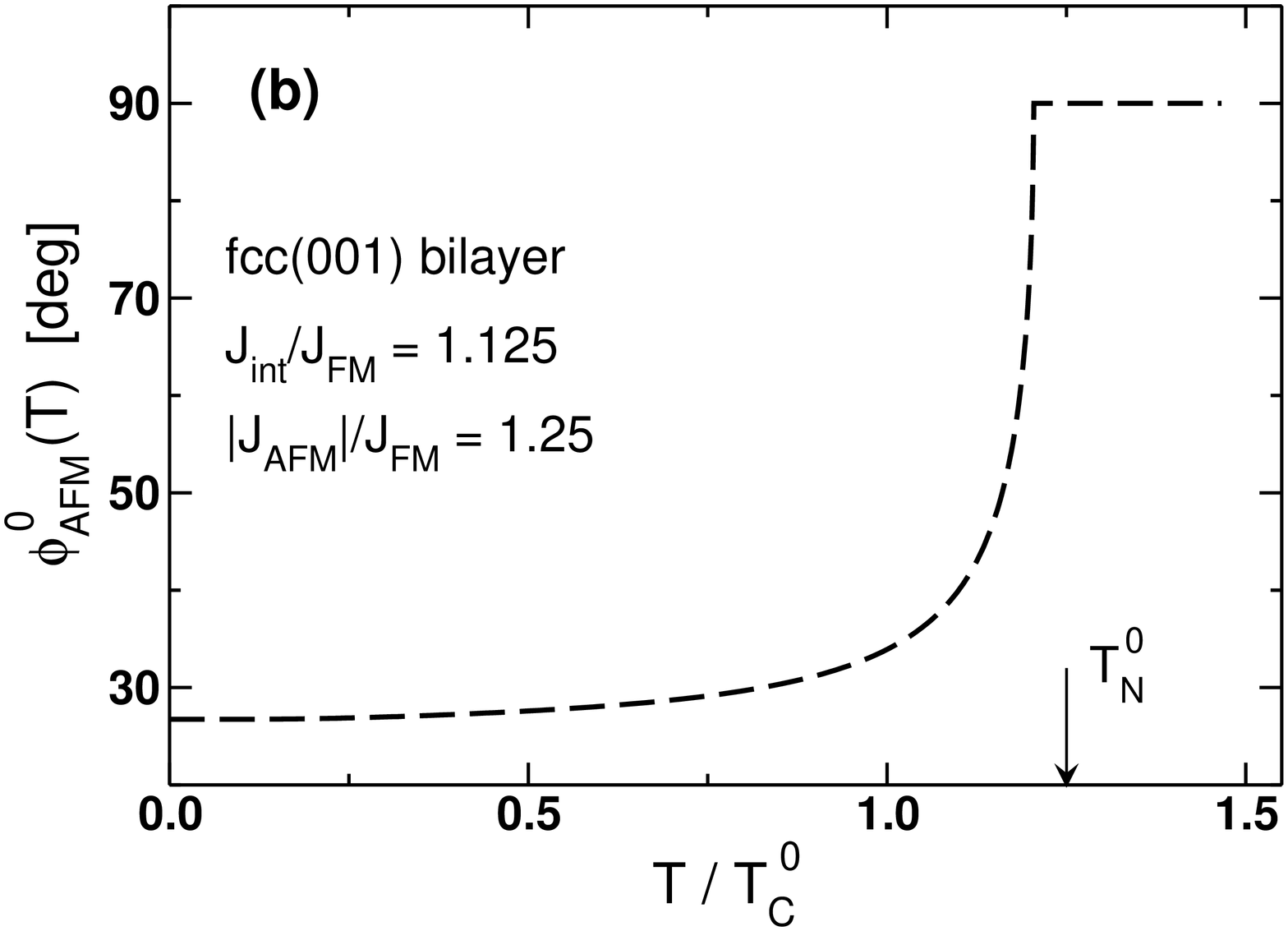} \\
\caption{ \label{fig7} 
  Same as Figure~\ref{fig6} for $J_\mathrm{int}/J_\mathrm{FM}=1.125$ and 
  $|J_\mathrm{AFM}|/J_\mathrm{FM}=1.25$, hence $T_N^0>T_C^0$. }
\end{figure}

\noindent b) $T_N^0>T_C^0$ and $J_\mathrm{int}>J_\mathrm{int}^*$. 
In effect this case is similar to the preceding one, i.e., with
increasing temperature the AFM spins rotate into the direction of
the FM.  However, although the FM exchange is weaker 
than the AFM exchange in this case, due to the strong interlayer 
coupling the FM layer \textit{dominates} the behavior of the AFM, and
results in an ordering temperature, cf.\ equation~(\ref{e14}),
even larger than $T_N^0$. The lack of a similar mechanism for
the AFM layer emphasizes the asymmetry of the two subsystems. 
Results are illustrated in Figure~\ref{fig7} for
different values of $J_\mathrm{int}$. 

\begin{figure}[t] 
\includegraphics[width=5.22cm,bb=45 25 500 760,angle=-90,clip]{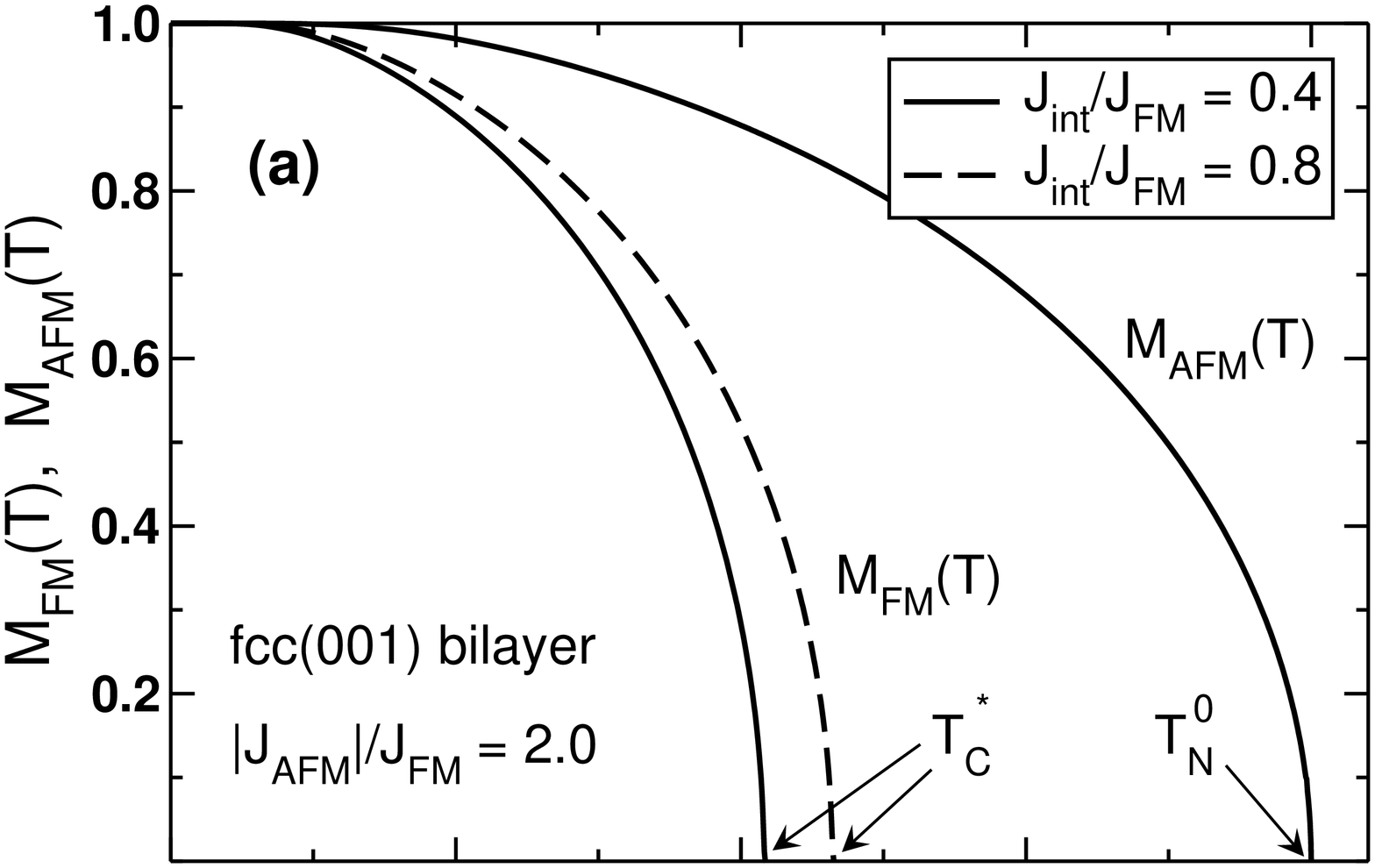} \\ 
\includegraphics[width=6cm,bb=45 25 575 760,angle=-90,clip]{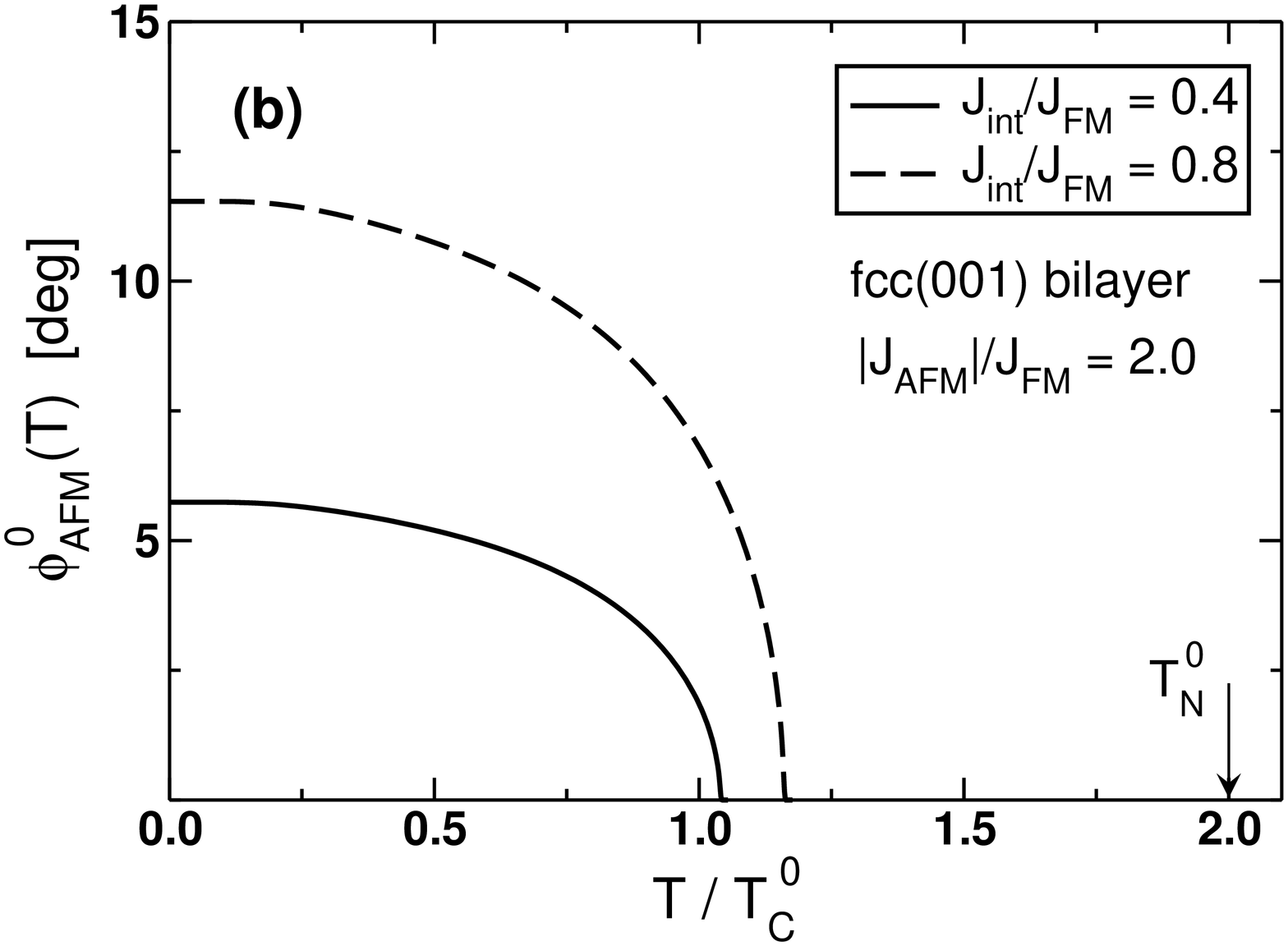} \\ 
\caption{\label{fig8} 
  Same as Figure~\ref{fig6} for $J_\mathrm{int}$ 
  as indicated, and $|J_\mathrm{AFM}|/J_\mathrm{FM}=2.0$, hence 
  $T_N^0>T_C^0$. For these systems two different
  critical temperatures $T_C^*$ and $T_N^0$ for the FM and AFM layers,
  respectively, are obtained. For $T\to T_C^*$ the AFM spins relax to
  the undisturbed AFM arrangement. }
\end{figure}

\noindent c) $T_N^0>T_C^0$ and $J_\mathrm{int}<J_\mathrm{int}^*$. 
The asymmetric behavior of the FM and AFM layers for the fcc(001) 
bilayer becomes even stronger. As before, the FM layer remains collinear. 
However, in this case the disturbance of the AFM layer and the 
angle $\phi_\mathrm{AFM}^0(T)$ \textit{decrease} with increasing 
temperature, as shown in Figure~\ref{fig8} for different values of 
$J_\mathrm{int}$.  
At the critical (Curie-) temperature $T_C^*>T_C^0$, given by
\begin{equation} \label{e16} 
T_C^*=\frac{S(S+1)}{3}\;\bigg[z_0\,J_\mathrm{FM}+
\frac{(z_1\,J_\mathrm{int})^2}{2\,z_0\,|J_\mathrm{AFM}|}\bigg] \;, 
\end{equation}  
the FM layer becomes paramagnetic, although in principle coupled to a
still ordered AFM layer.  However, no magnetization is induced in the
FM for $T>T_C^*$, since for $\phi_\mathrm{AFM}^0(T)=0$ the couplings
of an FM spin across the interface to the two AFM sublattices cancel
exactly, and since the scalar product of the interlayer exchange 
coupling, cf.\ equation~(\ref{e1}), vanishes for perpendicularly
oriented FM and AFM layers. 
The AFM layer becomes disordered at $T_C=T_N^0$, thus 
the bilayer ordering temperature is not given by equation~(\ref{e14}).
Evidently, in this case the interlayer exchange coupling
$J_\mathrm{int}$ is not strong enough to allow the FM layer to
dominate the AFM, like in the previous case. Hence, 
the coupled magnetic system has two critical temperatures. 
This behavior is present as long as $T_C^*$ is smaller than $T_N^0$. 
Equating $T_C^* =T_N^0$ yields the relation for the crossover 
interlayer coupling $J_\mathrm{int}^*$ given by equation~(\ref{e15}).

\section{Thicker Films} \label{sec:thick} 
\begin{figure}[t] 
\includegraphics[width=6cm,bb=50 25 575 760,angle=-90,clip]{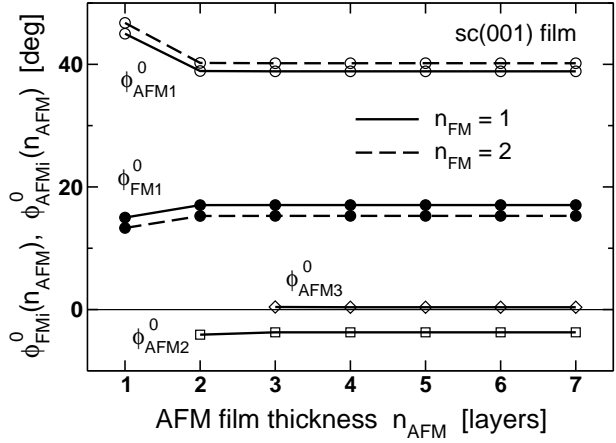} \\ 
\caption{\label{fig9} 
  Equilibrium angles $\phi_i^0$ of layers $i$ close to the 
  interface of an sc(001) FM-AFM film as function of the AFM film 
  thickness $n_\mathrm{AFM}$ at $T=0$. For the couplings we assume 
  $J_\mathrm{int}/J_\mathrm{FM}=4$ and 
  $|J_\mathrm{AFM}|/J_\mathrm{FM}=0.5$, and for the FM thickness 
  $n_\mathrm{FM}=1$ (solid lines) and $n_\mathrm{FM}=2$ (dashed 
  lines). Shown are $\phi_i^0$ of the FM interface layer (full 
  circles), the AFM interface layer (open circles), and the subsequent 
  two AFM layers (open squares and diamonds).} 
\end{figure}

In this Section we will present a number of results for coupled 
FM-AFM systems, where the individual FM and AFM films are thicker 
than just a monolayer. Evidently, the magnetizations 
$M_\mathrm{FMi}(T)$ and $M_\mathrm{AFMi}(T)$, and the sublattice 
canting angles $\phi_\mathrm{FMi}^0(T)$ and $\phi_\mathrm{AFMi}^0(T)$ 
will depend on the layer $i$. The deviation from the 
undisturbed magnetic arrangement, cf.\ Figures~\ref{fig1}a and 
\ref{fig3}a, is expected to be particularly pronounced for the 
layers close to the interface, whereas will vanish rapidly with
increasing distance from the interface. 
In Figure~\ref{fig9} the equilibrium angles are shown for an 
sc(001) lattice symmetry at $T=0$ as function of the AFM film 
thickness. For the FM film one and two layers are considered. The 
angles $\phi_\mathrm{i}^0$, particularly those close to the interface, 
saturate within two AFM layers, while thicker AFM films exhibit a 
weak oscillatory behavior of decreasing amplitude which cannot be 
observed on the scale of the figure. An alternating sign of 
$\phi_\mathrm{AFMi}^0$ is obtained for neighboring AFM layers. For 
distances from the interface larger than approximately three layers 
the AFM remains virtually undisturbed. A corresponding behavior is 
obtained by varying the FM film thickness. Similar results have been 
reported for instance in \cite{Fin04}. 

\begin{figure}[t] 
\includegraphics[width=5.15cm,bb=45 25 500 760,angle=-90,clip]{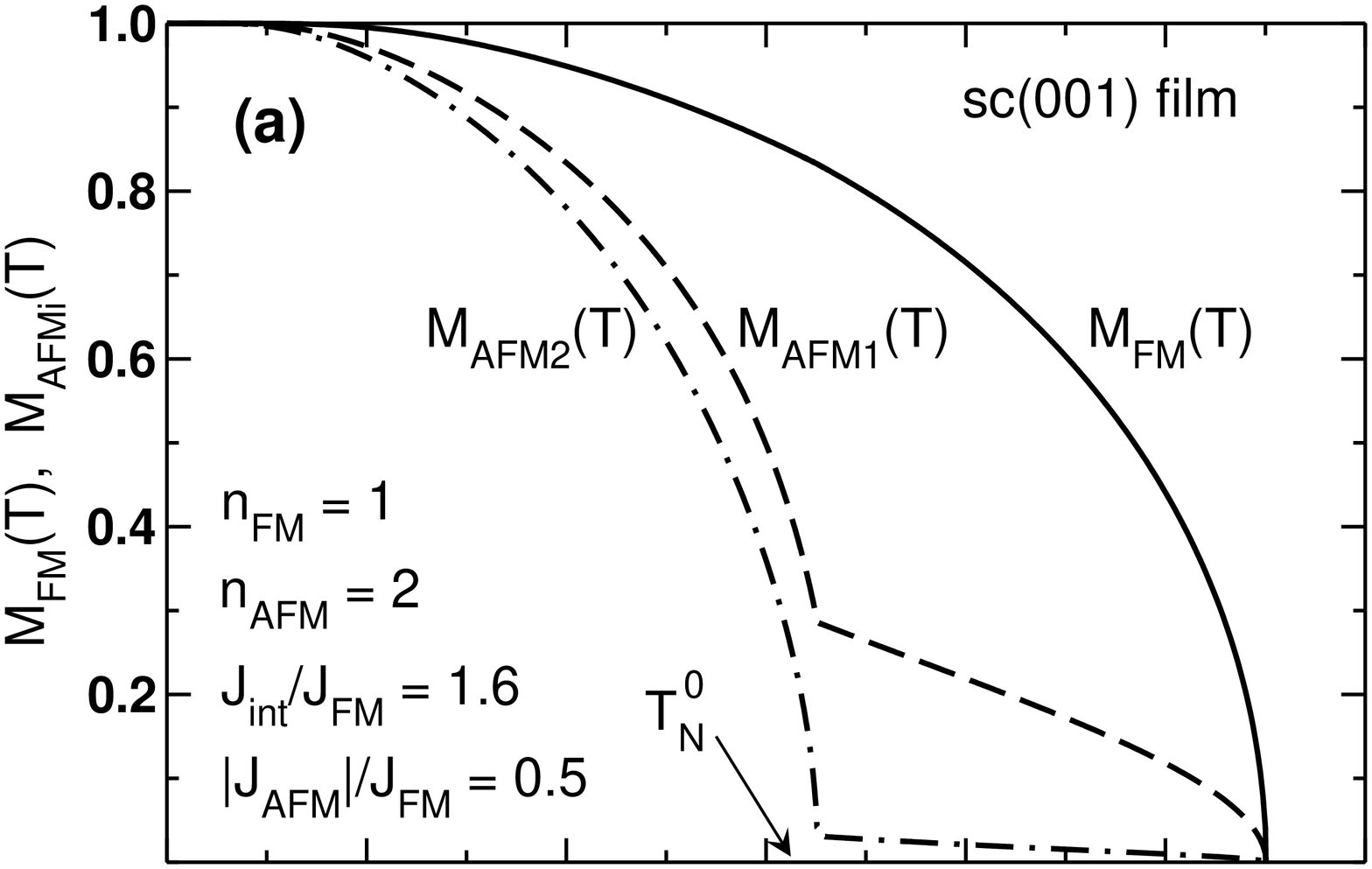} \\ 
\includegraphics[width=6cm,bb=45 25 575 760,angle=-90,clip]{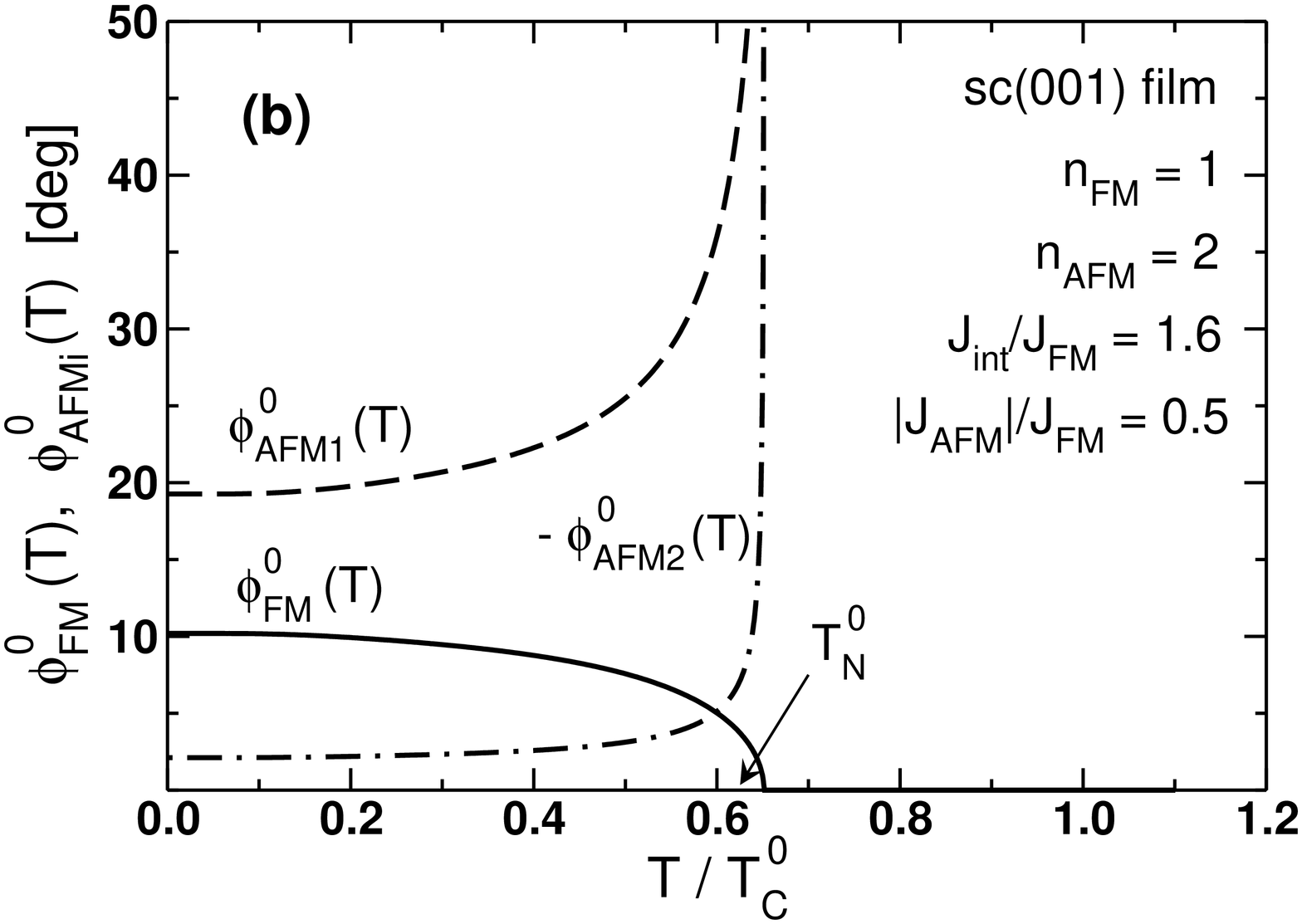} \\ 
\caption{\label{fig10}
  (a) Magnetizations $M_i(T)$ and (b) equilibrium angles $\phi_i^0(T)$ 
  for an sc(001) FM-AFM film as functions of the temperature
  $T/T_C^0$. We use $n_\mathrm{FM}=1$ and $n_\mathrm{AFM}=2$, moreover, 
  $J_\mathrm{int}/J_\mathrm{FM}=1.6$ and 
  $|J_\mathrm{AFM}|/J_\mathrm{FM}=0.5$. For convenience, for the
  second AFM layer we show $-\phi_\mathrm{AFM2}^0(T)$. }
\end{figure}

Moreover, we also investigate sc(001) FM-AFM systems with thicker AFM 
films at finite temperatures. As for the bilayer, and also for thicker 
films and for $T_N^0<T_C^0$, the AFM spins exhibit a rotation of the 
sublattice magnetization. With increasing temperature they turn into 
the direction of the FM film and become collinear above the sublattice 
reorientation temperature $T_R$, cf.\ Figure~\ref{fig10}.  The AFM 
magnetic arrangement for $T>T_R$ represents a `layered AFM structure' 
consisting of ferromagnetic layers with an alternating 
orientation for neigboring layers. All AFM layers become collinear at 
the same temperature, the variation of $\phi_\mathrm{AFMi}^0(T)$ is 
the steeper the larger the distance of layer $i$ from the interface. 
A similar behavior is also obtained for FM films thicker than a 
monolayer. In addition, for $T_N^0>T_C^0$ the behaviors of the FM and 
AFM subsystems are interchanged. Thus, the mentioned symmetry between 
FM and AFM films for the sc(001) symmetry, as calculated within MFA, is 
also present for thicker films. 

The discussion of the corresponding behavior of fcc(001) FM-AFM films 
requires some introductory remarks. Unlike FM films, and unlike
sc(001) AFM films, as calculated within MFA the N\'eel temperature 
$T_N(n_\mathrm{AFM})$ of an fcc(001) AFM film with an in-plane 
AFM order and with nearest neighbor exchange
interactions only does not increase with increasing thickness 
$n_\mathrm{AFM}$. Merely, a constant $T_N(n_\mathrm{AFM})$ given by 
the one of the monolayer ($n_\mathrm{AFM}=1$) results. 
Consequently, the same magnetizations $M_i(T)$, independent 
of the individual layer $i$, are obtained. Hence, the expression 
for the ordering temperature, cf.\ equation~(\ref{e14}), is not valid 
for an fcc(001) AFM bilayer. The reason is that for such a system 
with a collinear magnetization each layer is virtually decoupled from 
the rest. Only in the case of noncollinear magnetic order, as is
present e.g.\ close to the FM-AFM interface, a net coupling between
neighboring AFM layers results. 

\begin{figure}[t] 
\includegraphics[width=5.22cm,bb=50 25 500 760,angle=-90,clip]{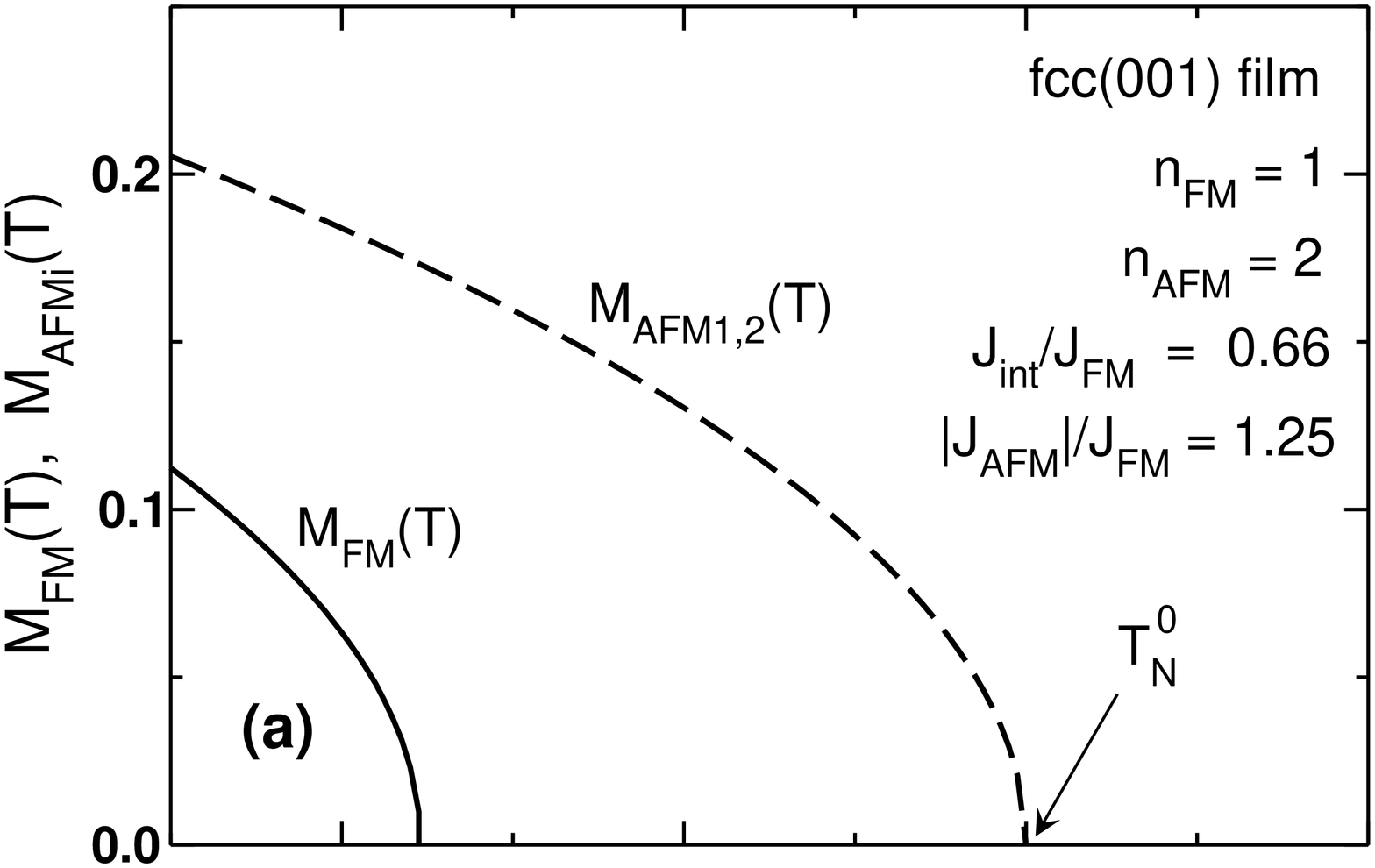} \\ 
\includegraphics[width=5.17cm,bb=55 25 500 760,angle=-90,clip]{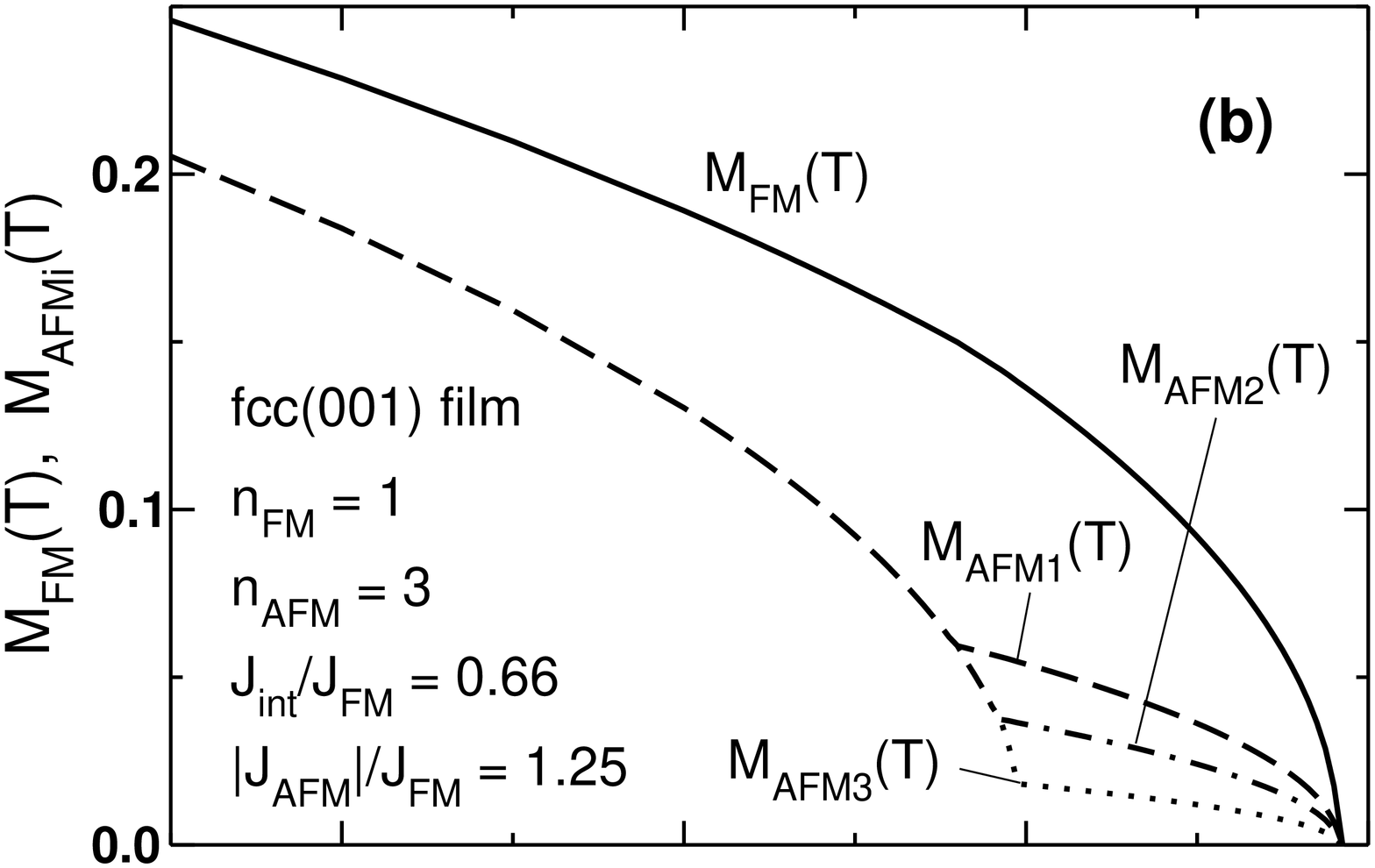} \\ 
\includegraphics[width=6cm,bb=55 25 570 760,angle=-90,clip]{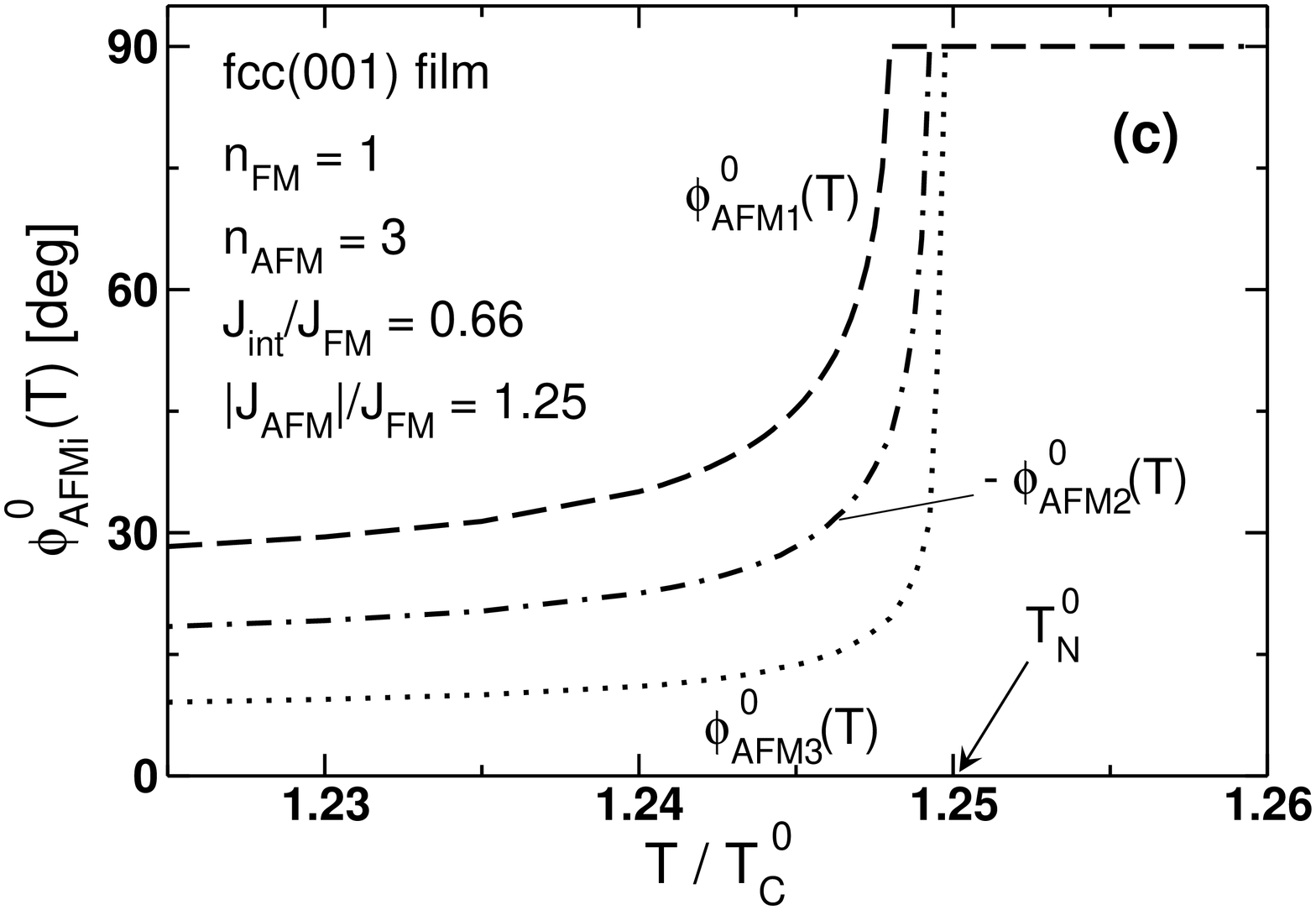} \\ 
\caption{\label{fig11} 
  Magnetizations $M_i(T)$ and equilibrium angles $\phi_i^0(T)$ for 
  fcc(001) FM-AFM films as functions of the temperature $T/T_C^0$. 
  We have assumed a single FM layer, $J_\mathrm{int}/J_\mathrm{FM}=0.66$, 
  and $|J_\mathrm{AFM}|/J_\mathrm{FM}=1.25$. (a) refers to 
  $n_\mathrm{AFM}=2$, (b) and (c) to $n_\mathrm{AFM}=3$. For the 
  second AFM layer $-\phi_\mathrm{AFM2}^0(T)$ is depicted. } 
\end{figure}

Keeping these features in mind we now discuss the 
finite-temperature properties of an FM-AFM system with an fcc(001) 
symmetry and for $n_\mathrm{AFM}>1$. As for the bilayer, all FM spins 
remain strictly collinear for all temperatures. In Figure~\ref{fig11} 
the magnetizations $M_i(T)$ and angles $\phi_i^0(T)$ 
for FM-AFM films with $n_\mathrm{AFM}=2$ and $n_\mathrm{AFM}=3$ 
close to their critical temperatures $T_C$ 
is presented. The FM film thickness $n_\mathrm{FM}=1$ 
and the coupling constants are the same for both cases and are chosen 
in such a way that $T_N^0>T_C^0$. The case $n_\mathrm{AFM}=2$ 
corresponds to the situation shown in Figure~\ref{fig8}. 
The FM layer becomes paramagnetic above the critical temperature 
$T_C^*$ in the presence of a still ordered AFM film. Thus, two 
critical temperatures can occur also for thicker FM-AFM films. The 
magnetizations $M_i(T)$ of the two AFM layers are identical to each 
other over the whole temperature range and vanish at 
$T_N^0(n_\mathrm{AFM}=1)$, cf.\ Figure~\ref{fig11}a. In contrast, the 
equilibrium angles $\phi_\mathrm{AFMi}^0(T)$ are different for both 
AFM layers, and approach $\phi_\mathrm{AFMi}^0(T)\to0$ for 
$T\to T_C^*$ (not shown). 

For the applied coupling constants this behavior changes 
drastically if three AFM layers are considered. Although still 
$T_C^0<T_N^0$, the FM layer now dominates and causes a similar 
behavior as shown in Figure~\ref{fig7} for an FM-AFM bilayer with a 
strong interlayer coupling. As can be seen from Figure~\ref{fig11}c, 
the angles $\phi_\mathrm{AFMi}^0(T)$ of the AFM layers increase with 
increasing temperature. The AFM spins eventually become collinear with 
respect to the FM, with an alternating magnetic orientation for 
neighboring AFM layers. In contrast to the sc(001) film 
shown in Figure~\ref{fig10}, the sublattice reorientation temperature 
$T_{R,i}$ is now layer dependent and increases as the distance of the 
layer $i$ from the interface becomes larger. Moreover, as long as the 
AFM layers maintain a noncollinear structure, the magnetizations 
$M_i(T)$ are identical and independent of the layer index. Only for 
temperatures $T>T_{R,i}$ the $M_i(T)$ differ from each other, 
and vanish together with the magnetization of the FM film at the 
common ordering temperature $T_C$ of the total FM-AFM system, cf.\ 
Figure~\ref{fig11}b. 

The different behavior of the AFM magnetizations in coupled sc(001) 
and fcc(001) FM-AFM films can be understood as follows. For the former 
symmetry the magnetic structures of all FM and AFM layers are disturbed 
for $T=0$, and become collinear at the same temperature. On the other 
hand, for fcc(001) films the FM layers always remain collinear 
($\phi_\mathrm{FMi}^0(T)=0$). Consider the situation depicted in 
Figures~\ref{fig11}b,c. If, e.g., the spins of the AFM interface layer 
(AFM1) turn into the direction of the FM, the remaining AFM layers 
virtually experience an ordered FM film with an increased thickness. 
As before, the remaining AFM layers can maintain a noncollinear 
magnetic arrangement, and there is no need for all layers 
to become simultaneously collinear.  In addition, we note that the AFM 
films above the sublattice reorientation temperatures exhibit, for both 
AFM thicknesses shown in Figure~\ref{fig11}, a collinear structure. 
Nevertheless they behave differently since for $n_\mathrm{AFM}=2$, 
Figure~\ref{fig11}a, the magnetic structure refers to
an `in-plane AFM' for $T>T_C^*$, and for $n_\mathrm{AFM}=3$ and $T>T_{R,i}$, 
Figure~\ref{fig11}b,c, to a `layered AFM structure'. In the latter case 
the AFM magnetizations $M_i(T)$ are layer dependent, and the 
corresponding ordering temperature depends on the AFM film thickness. 

\section{Conclusion} \label{sec:concl} 
In this study 
we investigated how the magnetic structure rearranges in the vicinity 
of the interface between a ferromagnet and an antiferromagnet. Thin 
film systems with sc(001) and fcc(001) symmetries have been solved for 
both zero and finite temperatures within the framework of a mean field 
approximation.  A variety of configurations was obtained, and the 
underlying physics has been discussed. In contrast with previous work 
\cite{Koo97,HiM86,CrC01,JeD02}, these properties are mainly 
determined by the isotropic exchange interactions. The consideration 
of a particularly simple bilayer system, and the application of an MFA 
at finite temperatures, allows us to derive \textit{analytical} 
expressions for various quantities. These serve as estimates of the 
magnetic behavior for more realistic coupled FM-AFM systems having 
thicker FM and AFM films. 

We emphasize the different behavior of the  sc(001) and fcc(001) 
lattice symmetries. In particular, a canting of the sublattice 
magnetizations of both FM and AFM layers is obtained for the former
case, whereas for the latter only the AFM layer is
disturbed. Moreover, if the bare Curie temperature $T_C^0$ of the FM
film is larger than the bare N\'eel temperature $T_N^0$ of the AFM film, 
the AFM spins become collinear with respect to the FM system above the 
sublattice reorientation temperature $T_R$ for both investigated
symmetries. For $T>T_R$ the AFM film assumes a `layered AFM
structure'. For an sc(001) lattice this 
reorientation happens simultaneously for all layers at the same 
temperature. For $T_N^0>T_C^0$ a corresponding behavior 
with an interchanged role of the FM and AFM films results, which
within MFA is perfectly symmetric to the case $T_N^0<T_C^0$. 
In contrast, such a symmetry between FM and AFM is not present for the 
fcc(001) lattice. Merely, the FM spins always remain strictly
collinear. The different AFM layers turn into the direction of the FM
at different sublattice reorientation temperatures. 

Moreover, the possibility of two critical temperatures is pointed 
out, as derived for fcc(001) FM-AFM bilayers for $T_N^0>T_C^0$ and 
$J_\mathrm{int}<J_\mathrm{int}^*$. In this case the FM film becomes 
paramagnetic at temperatures $T_C^*$ where the AFM film is still 
magnetically ordered. The presence of two different $T_C$'s in 
magnetic systems is well known, for instance, for two coupled 
semi-infinite ferromagnets. Similarly, if a magnetic film 
with a strong exchange interaction is deposited on 
a bulk ferromagnet, two different ordering temperatures may 
exist \cite{KTS83}. In contrast, to our knowledge two critical 
temperatures for coupled magnetic films with finite thicknesses 
have not been reported previously. 

However, the existence of two $T_C$'s is expected to be fragile. In
fact, small deviations from the fcc(001) symmetry, for example in 
presence of disorder near the interface, could destroy the lower one. 
The reason is that in this case the couplings 
of the two AFM sublattices across the interface do not cancel exactly, 
and a magnetization will be induced in the FM for $T>T_C^*$. 
In general, we note that in real FM-AFM interfaces disorder is always
present, like step, vacancies, interdiffusion, etc. In this case the
lateral periodicity of the magnetic structure as sketched in
Figures~\ref{fig1}b and \ref{fig3}b will vanish with increasing 
degree of disorder. The presented results are obtained for fully
ordered interfaces and thus will serve as starting points to
investigate the role of disorder at FM-AFM interfaces. For example,
the resulting magnetic arrangement can be a mixture of the two
extremal cases represented by the sc(001) and fcc(001) stackings. 
For a strong disorder compensated and noncompensated interfaces can
no longer be distinguished \cite{Fin04}. Moreover, as mentioned in the
Introduction, the consideration of disorder seems to be essential to 
explain the exchange bias effect \cite{MNL98}. 

As noted in Section~\ref{sec:theory}, we have chosen anisotropy 
easy axes of the FM and AFM films which support a perdendicular 
magnetic arrangement of both subsystems. 
Anisotro\-pies with different symmetries and arbitrary directions of 
the easy axes can be considered as well. In that case the magnetic 
structure and the (sublattice) spin rotation will also depend on 
the anisotropies. In presence of disorder the anisotropy easy axes 
will be site-dependent which, if the anisotropy is sufficiently
strong, can disturb the lateral periodicity of the magnetic
arrangements sketched in Figures~\ref{fig1}b and \ref{fig3}b. 

In this connection we like to point out an important difference with 
our prior work \cite{JeD02}, which also dealt with coupled FM-AFM 
films.  There the magnetization of the FM undergoes a \textit{full 
spin reorientation transition} (SRT) as a function of temperature, 
i.e., the net magnetization of each layer changes its direction 
whereas its magnitude stays approximately constant. To exhibit such an 
SRT a significant anisotropy in the FM must be present, eventually 
competing with the interlayer exchange. In contrast, in the present 
study both sublattices in every layer exhibit a magnetic reorientation, 
with opposite sense of the rotations. The directions of the net 
layer magnetizations remain constant and do not show an SRT, whereas 
their magnitudes vary considerably.
These differences should become apparent in possible experimental
realizations, e.g., within an element-specific X-ray magnetic
linear or circular dichroism (XMLD, XMCD) measurement \cite{SLA04}. 
Whether a full SRT like in \cite{JeD02}, or whether the magnetic 
arrangement as described in the present study dominates, depends on 
the actual FM-AFM system under consideration.  

Finally, we like to discuss the influence of collective magnetic
excitations (spin waves). As is well known, for 2D magnetic systems 
these excitations play a very important role, which however are 
neglected in the MFA used in this study. It is therefore important to 
apply improved methods which take into account collective excitations, 
for example, within a many-body Green's function theory (GFT) 
\cite{Tya67}. FM-AFM bilayer and multilayer systems have been
investigated previously by this method, considering a collinear 
magnetization \cite{Die89}. In \cite{SuS98} the collective excitations 
were discussed to be a possible source for the exchange bias effect. 
The GFT has recently been generalized \cite{JKW03} to take into
account several nonvanishing components of the magnetization, hence 
allowing the investigation of noncollinear magnetic strucures. 
To avoid the catastrophe of the Mermin-Wagner-theorem \cite{MeW66}
magnetic anisotropies must be incorporated explicitly.  Analytical 
results, which can be drawn from the much simpler MFA, may not be 
obtained from such improved theoretical approaches. Also, it has been 
shown that MFA yields at least \textit{qualitatively} correct results 
for anisotropic magnetic thin films, although quantitatively it 
strongly overestimates the ordering temperatures. Preliminary results 
calculated with GFT show that the main properties obtained in the 
present study are supported. In particular, this is valid for the 
sublattice magnetic 
reorientation, and the distortion of both FM and AFM layers in case of 
sc(001) FM-AFM films. On the other hand, the exact symmetry 
between the FM and AFM layers for the sc(001) system turns out to be an 
artifact of the MFA. The reason is that the spin wave dispersion 
relations for an FM and an AFM differ qualitatively, as do the 
respective ordering temperatures even for the same strengths of the 
exchange couplings. However, MFA incorrectly yields the same ordering 
temperatures. Further investigations using GFT are underway. 

PJJ acknowledges support by the Deutsche  
For\-schungs\-gemeinschaft through SFB 290, TP A1. HD and MK acknowledge
support by ECOS. MK was also supported by Fundaci\'on Andes and by
FONDECYT, Chile, under grant No.\ 1030957.  


\vspace{0.5cm}

\begin{thebibliography}{99}

\bibitem{Zuc73} M. J. Zuckermann, Solid State Comm. \textbf{12}, 745 (1973).  

\bibitem{KiZ73} M. Kiwi and M. J. Zuckermann, Proc.\ of the 19th 
  Conf.\ on Magnetism and Magnetic Materials \textbf{18}, 347 (1973); 
  B. N. Cox, R. A. Tahir-Kheli, and R. J. Elliott, Phys.
  Rev. B \textbf{20}, 2864 (1979); 
  G. J. Mata, E. Pestana, and M. Kiwi, Phys. Rev. B \textbf{26}, 3841 (1982);
  D. Altbir, M. Kiwi, G. Mart{\'\i}nez, and M. J. Zuckermann,
  Phys. Rev. B \textbf{40}, 6963 (1989); 
  J. Tersoff and L. M. Falicov, Phys. Rev. B \textbf{25}, 
  R2959 (1982); \textit{ibid.} \textbf{26}, 459, 6186 (1982);  
  J. Mathon, M. Villeret, R. B. Muniz, J. d'Albuquerque e Castro, 
  and D. M. Edwards, Phys. Rev. Lett. \textbf{74}, 3696 (1995); 
  H. Hasegawa and F. Herman, Phys. Rev. B \textbf{38},
  4863 (1988).

\bibitem{NoS99} For recent reviews see, e.g.: J. Nogu\'es and I. K. 
  Schuller, J. Magn. Magn. Mater. \textbf{192}, 203 (1999); 
  M. Kiwi, \textit{ibid.} \textbf{234/3}, 584 (2001);
  R. L. Stamps, J. Phys. D: Appl. Phys. \textbf{33}, R247 (2000). 

\bibitem{Nee67} L. N\'eel, Ann. Phys. (Paris) \textbf{2}, 61 (1967).
  
\bibitem{MNL98} T. J. Moran, J. Nogu\'es, D. Lederman, and I. K.
  Schuller, Appl. Phys. Lett. \textbf{72}, 617 (1998); 
  M. Kiwi, J. J. Mej{\'\i}a-L\'opez, R. D. Portugal, and R.
  Ram{\'\i}rez, \textit{ibid.} \textbf{75}, 3995 (1999); 
  T. C. Schulthess and W. H. Butler, Phys. Rev. Lett.
  \textbf{81}, 4516 (1998); 
  K. Takano, R. H.  Kodama, A. E. Berkowitz, W. Cao, and
  G. Thomas, \textit{ibid.} \textbf{79}, 1130 (1997);  
  P. Milt\'enyi, M. Gierlings, J. Keller, B. Beschoten,
  G. G\"untherodt, U. Nowak, and K. D. Usadel, \textit{ibid.} 
  \textbf{84}, 4224 (2000).

\bibitem{Koo97} N. C. Koon, Phys. Rev. Lett. \textbf{78}, 4865 (1997).

\bibitem{HiM86} L. L. Hinchey and D. L. Mills, Phys. Rev. B \textbf{33},
  3329 (1986).

\bibitem{CrC01} L. L. Hinchey and D. L. Mills, Phys. Rev. B \textbf{34},
  1689 (1986); 
  N. Cramer and R. E. Camley,  \textit{ibid.} \textbf{63}, 060404(R) (2001).

\bibitem{JeD02} P. J. Jensen and H. Dreyss\'e, Phys. Rev. B 
  \textbf{66}, 220407(R) (2002). 

\bibitem{MeW66} N. D. Mermin and H. Wagner, Phys. Rev. Lett. 
  \textbf{17}, 1133 (1966). 

\bibitem{hexagonal} Hexagonal (111) lattice faces can also be
  considered appropriately. Since AFM (111) layers break up into three
  sublattices, the analysis will be slightly more complicated. 
 
\bibitem{dipol} Note that for an AFM monolayer the dipole interaction 
  preferres a perpendicular magnetization, however with a much smaller 
  energy gain than the demagnetizing energy of an FM layer 
  (P. J. Jensen, Ann. Physik (Leipzig) \textbf{6}, 317 (1997);  
  P. Politi and M. G. Pini, Phys. Rev. B \textbf{66}, 214414 (2002)). 

\bibitem{Fin04} M. Finazzi, Phys. Rev. B \textbf{69}, 064405 (2004). 

\bibitem{KTS83} T. Kaneyoshi, I. Tamura, and E. F. Sarmento, 
  Phys. Rev. B \textbf{28}, 6491 (1983); J. M. Sanchez and 
  J. L. Mor\'an-L\'opez, in \textit{Magnetic Properties of
  Low-dimensional systems}, edited by L. M. Falicov and 
  J. L. Mor\'an-L\'opez, Springer Verlag Berlin (1986), p. 114;
  P. J. Jensen, H. Dreyss\'e, and K. H. Bennemann, 
  Europhys. Lett. \textbf{18}, 463 (1992).

\bibitem{SLA04} A. Scholl, M. Liberati, E. Arenholz, H. Ohldag, and 
  J. St\"ohr, Phys. Rev. Lett. \textbf{92}, 247201 (2004). 

\bibitem{Tya67} S. V. Tyablikov, \textit{Methods in the quantum theory
  of magnetism} (Plenum Press, New York, 1967); W. Nolting,
  \textit{Quantentheorie des Magnetismus} (B. G. Teubner, Stuttgart,
  1986), Vol.2.

\bibitem{Die89} H. T. Diep, Phys. Rev. B \textbf{40}, 4818 (1989); 
  A. Moschel, K. D. Usadel, and A. Hucht, Phys. Rev. B \textbf{47}, 
  8676 (1993); A. Moschel and K. D. Usadel, Phys. Rev. B \textbf{48}, 
  13991 (1993). 

\bibitem{SuS98} H. Suhl and I. K. Schuller, Phys. Rev. B \textbf{58}, 
  258 (1998). 

\bibitem{JKW03} P. J. Jensen, S. Knappmann, W. Wulfhekel, and
  H. P. Oepen, Phys. Rev. B \textbf{67}, 184417 (2003); 
  P. Fr\"obrich and P. J. Kuntz, Europ. Phys. J. B \textbf{32}, 445
  (2003). 

\end{thebibliography}
\end{document}